\pdfoutput=1

\documentclass[11pt,twoside,a4paper,cmspaper,final,collab]{cms-tdr}
\def\svnVersion{37723:38250M}\def\cmsCernNo{2011-001}\def\cmsCernDate{2011/02/10}\def\cmsMessage{Submitted to Physical Review C}

\begin{document}\cmsNoteHeader{HIN-10-004}

\hyphenation{had-ron-i-za-tion}
\hyphenation{cal-or-i-me-ter}
\hyphenation{de-vices}
\RCS$Revision: 37951 $
\RCS$HeadURL: svn+ssh://alverson@svn.cern.ch/reps/tdr2/papers/HIN-10-004/trunk/HIN-10-004.tex $
\RCS$Id: HIN-10-004.tex 37951 2011-02-09 01:42:18Z alverson $
%
%
%

\providecommand {\etal}{\mbox{et al.}\xspace} 
\providecommand {\ie}{\mbox{i.e.}\xspace}     
\providecommand {\eg}{\mbox{e.g.}\xspace}     
\providecommand {\etc}{\mbox{etc.}\xspace}     
\providecommand {\vs}{\mbox{\sl vs.}\xspace}      
\providecommand {\mdash}{\ensuremath{\mathrm{-}}} 

\providecommand {\Lone}{Level-1\xspace} 
\providecommand {\Ltwo}{Level-2\xspace}
\providecommand {\Lthree}{Level-3\xspace}

\providecommand{\ACERMC} {\textsc{AcerMC}\xspace}
\providecommand{\ALPGEN} {{\textsc{alpgen}}\xspace}
\providecommand{\CHARYBDIS} {{\textsc{charybdis}}\xspace}
\providecommand{\CMKIN} {\textsc{cmkin}\xspace}
\providecommand{\CMSIM} {{\textsc{cmsim}}\xspace}
\providecommand{\CMSSW} {{\textsc{cmssw}}\xspace}
\providecommand{\COBRA} {{\textsc{cobra}}\xspace}
\providecommand{\COCOA} {{\textsc{cocoa}}\xspace}
\providecommand{\COMPHEP} {\textsc{CompHEP}\xspace}
\providecommand{\EVTGEN} {{\textsc{evtgen}}\xspace}
\providecommand{\FAMOS} {{\textsc{famos}}\xspace}
\providecommand{\GARCON} {\textsc{garcon}\xspace}
\providecommand{\GARFIELD} {{\textsc{garfield}}\xspace}
\providecommand{\GEANE} {{\textsc{geane}}\xspace}
\providecommand{\GEANTfour} {{\textsc{geant4}}\xspace}
\providecommand{\GEANTthree} {{\textsc{geant3}}\xspace}
\providecommand{\GEANT} {{\textsc{geant}}\xspace}
\providecommand{\HDECAY} {\textsc{hdecay}\xspace}
\providecommand{\HERWIG} {{\textsc{herwig}}\xspace}
\providecommand{\HIGLU} {{\textsc{higlu}}\xspace}
\providecommand{\HIJING} {{\textsc{hijing}}\xspace}
\providecommand{\IGUANA} {\textsc{iguana}\xspace}
\providecommand{\ISAJET} {{\textsc{isajet}}\xspace}
\providecommand{\ISAPYTHIA} {{\textsc{isapythia}}\xspace}
\providecommand{\ISASUGRA} {{\textsc{isasugra}}\xspace}
\providecommand{\ISASUSY} {{\textsc{isasusy}}\xspace}
\providecommand{\ISAWIG} {{\textsc{isawig}}\xspace}
\providecommand{\MADGRAPH} {\textsc{MadGraph}\xspace}
\providecommand{\MCATNLO} {\textsc{mc@nlo}\xspace}
\providecommand{\MCFM} {\textsc{mcfm}\xspace}
\providecommand{\MILLEPEDE} {{\textsc{millepede}}\xspace}
\providecommand{\ORCA} {{\textsc{orca}}\xspace}
\providecommand{\OSCAR} {{\textsc{oscar}}\xspace}
\providecommand{\PHOTOS} {\textsc{photos}\xspace}
\providecommand{\PROSPINO} {\textsc{prospino}\xspace}
\providecommand{\PYTHIA} {{\textsc{pythia}}\xspace}
\providecommand{\SHERPA} {{\textsc{sherpa}}\xspace}
\providecommand{\TAUOLA} {\textsc{tauola}\xspace}
\providecommand{\TOPREX} {\textsc{TopReX}\xspace}
\providecommand{\XDAQ} {{\textsc{xdaq}}\xspace}

\providecommand {\DZERO}{D\O\xspace}     


\providecommand{\de}{\ensuremath{^\circ}}
\providecommand{\ten}[1]{\ensuremath{\times \text{10}^\text{#1}}}
\providecommand{\unit}[1]{\ensuremath{\text{\,#1}}\xspace}
\providecommand{\mum}{\ensuremath{\,\mu\text{m}}\xspace}
\providecommand{\micron}{\ensuremath{\,\mu\text{m}}\xspace}
\providecommand{\cm}{\ensuremath{\,\text{cm}}\xspace}
\providecommand{\mm}{\ensuremath{\,\text{mm}}\xspace}
\providecommand{\mus}{\ensuremath{\,\mu\text{s}}\xspace}
\providecommand{\keV}{\ensuremath{\,\text{ke\hspace{-.08em}V}}\xspace}
\providecommand{\MeV}{\ensuremath{\,\text{Me\hspace{-.08em}V}}\xspace}
\providecommand{\GeV}{\ensuremath{\,\text{Ge\hspace{-.08em}V}}\xspace}
\providecommand{\gev}{\GeV}
\providecommand{\TeV}{\ensuremath{\,\text{Te\hspace{-.08em}V}}\xspace}
\providecommand{\PeV}{\ensuremath{\,\text{Pe\hspace{-.08em}V}}\xspace}
\providecommand{\keVc}{\ensuremath{{\,\text{ke\hspace{-.08em}V\hspace{-0.16em}/\hspace{-0.08em}}c}}\xspace}
\providecommand{\MeVc}{\ensuremath{{\,\text{Me\hspace{-.08em}V\hspace{-0.16em}/\hspace{-0.08em}}c}}\xspace}
\providecommand{\GeVc}{\ensuremath{{\,\text{Ge\hspace{-.08em}V\hspace{-0.16em}/\hspace{-0.08em}}c}}\xspace}
\providecommand{\TeVc}{\ensuremath{{\,\text{Te\hspace{-.08em}V\hspace{-0.16em}/\hspace{-0.08em}}c}}\xspace}
\providecommand{\keVcc}{\ensuremath{{\,\text{ke\hspace{-.08em}V\hspace{-0.16em}/\hspace{-0.08em}}c^\text{2}}}\xspace}
\providecommand{\MeVcc}{\ensuremath{{\,\text{Me\hspace{-.08em}V\hspace{-0.16em}/\hspace{-0.08em}}c^\text{2}}}\xspace}
\providecommand{\GeVcc}{\ensuremath{{\,\text{Ge\hspace{-.08em}V\hspace{-0.16em}/\hspace{-0.08em}}c^\text{2}}}\xspace}
\providecommand{\TeVcc}{\ensuremath{{\,\text{Te\hspace{-.08em}V\hspace{-0.16em}/\hspace{-0.08em}}c^\text{2}}}\xspace}

\providecommand{\pbinv} {\mbox{\ensuremath{\,\text{pb}^\text{$-$1}}}\xspace}
\providecommand{\fbinv} {\mbox{\ensuremath{\,\text{fb}^\text{$-$1}}}\xspace}
\providecommand{\nbinv} {\mbox{\ensuremath{\,\text{nb}^\text{$-$1}}}\xspace}
\providecommand{\percms}{\ensuremath{\,\text{cm}^\text{$-$2}\,\text{s}^\text{$-$1}}\xspace}
\providecommand{\lumi}{\ensuremath{\mathcal{L}}\xspace}
\providecommand{\Lumi}{\ensuremath{\mathcal{L}}\xspace}
%
\providecommand{\LvLow}  {\ensuremath{\mathcal{L}=\text{10}^\text{32}\,\text{cm}^\text{$-$2}\,\text{s}^\text{$-$1}}\xspace}
\providecommand{\LLow}   {\ensuremath{\mathcal{L}=\text{10}^\text{33}\,\text{cm}^\text{$-$2}\,\text{s}^\text{$-$1}}\xspace}
\providecommand{\lowlumi}{\ensuremath{\mathcal{L}=\text{2}\times \text{10}^\text{33}\,\text{cm}^\text{$-$2}\,\text{s}^\text{$-$1}}\xspace}
\providecommand{\LMed}   {\ensuremath{\mathcal{L}=\text{2}\times \text{10}^\text{33}\,\text{cm}^\text{$-$2}\,\text{s}^\text{$-$1}}\xspace}
\providecommand{\LHigh}  {\ensuremath{\mathcal{L}=\text{10}^\text{34}\,\text{cm}^\text{$-$2}\,\text{s}^\text{$-$1}}\xspace}
\providecommand{\hilumi} {\ensuremath{\mathcal{L}=\text{10}^\text{34}\,\text{cm}^\text{$-$2}\,\text{s}^\text{$-$1}}\xspace}


\providecommand{\PT}{\ensuremath{p_{\mathrm{T}}}\xspace}
\providecommand{\pt}{\ensuremath{p_{\mathrm{T}}}\xspace}
\providecommand{\ET}{\ensuremath{E_{\mathrm{T}}}\xspace}
\providecommand{\HT}{\ensuremath{H_{\mathrm{T}}}\xspace}
\providecommand{\et}{\ensuremath{E_{\mathrm{T}}}\xspace}
\providecommand{\Em}{\ensuremath{E\hspace{-0.6em}/}\xspace}
\providecommand{\Pm}{\ensuremath{p\hspace{-0.5em}/}\xspace}
\providecommand{\PTm}{\ensuremath{{p}_\mathrm{T}\hspace{-1.02em}/}\xspace}
\providecommand{\PTslash}{\ensuremath{{p}_\mathrm{T}\hspace{-1.02em}/}\xspace}
\providecommand{\ETm}{\ensuremath{E_{\mathrm{T}}^{\text{miss}}}\xspace}
\providecommand{\MET}{\ETm}
\providecommand{\ETmiss}{\ETm}
\providecommand{\ETslash}{\ensuremath{E_{\mathrm{T}}\hspace{-1.1em}/}\xspace}
\providecommand{\VEtmiss}{\ensuremath{{\vec E}_{\mathrm{T}}^{\text{miss}}}\xspace}

\providecommand{\dd}[2]{\ensuremath{\frac{\mathrm{d} #1}{\mathrm{d} #2}}}
\providecommand{\ddinline}[2]{\ensuremath{\mathrm{d} #1/\mathrm{d} #2}}

\ifthenelse{\boolean{cms@italic}}{\providecommand{\cmsSymbolFace}{\relax}}{\providecommand{\cmsSymbolFace}{\mathrm}}

\providecommand{\zp}{\ensuremath{\cmsSymbolFace{Z}^\prime}\xspace}
\providecommand{\JPsi}{\ensuremath{\cmsSymbolFace{J}\hspace{-.08em}/\hspace{-.14em}\psi}\xspace}
\providecommand{\Z}{\ensuremath{\cmsSymbolFace{Z}}\xspace}
\providecommand{\ttbar}{\ensuremath{\cmsSymbolFace{t}\overline{\cmsSymbolFace{t}}}\xspace}

\providecommand{\cPgn}{\ensuremath{\nu}\xspace}
\providecommand{\cPJgy}{\JPsi}
\providecommand{\cPZ}{\Z}
\providecommand{\cPZpr}{\zp}
\providecommand{\cPqb}{\ensuremath{\cmsSymbolFace{b}}\xspace} 
\providecommand{\cPqt}{\ensuremath{\cmsSymbolFace{t}}\xspace} 
\providecommand{\cPqc}{\ensuremath{\cmsSymbolFace{c}}\xspace} 
\providecommand{\cPaqb}{\ensuremath{\overline{\cmsSymbolFace{b}}\xspace}} 
\providecommand{\cPaqt}{\ensuremath{\overline{\cmsSymbolFace{t}}}\xspace} 
\providecommand{\cPaqc}{\ensuremath{\overline{\cmsSymbolFace{c}}}\xspace} 


\providecommand{\AFB}{\ensuremath{A_\text{FB}}\xspace}
\providecommand{\wangle}{\ensuremath{\sin^{2}\theta_{\text{eff}}^\text{lept}(M^2_\Z)}\xspace}
\providecommand{\stat}{\ensuremath{\,\text{(stat.)}}\xspace}
\providecommand{\syst}{\ensuremath{\,\text{(syst.)}}\xspace}
\providecommand{\kt}{\ensuremath{k_{\mathrm{T}}}\xspace}

\providecommand{\BC}{\ensuremath{\mathrm{B_{c}}}\xspace}
\providecommand{\bbarc}{\ensuremath{\mathrm{\overline{b}c}}\xspace}
\providecommand{\bbbar}{\ensuremath{\mathrm{b\overline{b}}}\xspace}
\providecommand{\ccbar}{\ensuremath{\mathrm{c\overline{c}}}\xspace}
\providecommand{\bspsiphi}{\ensuremath{\mathrm{B_s} \to \JPsi\, \phi}\xspace}
\providecommand{\EE}{\ensuremath{\mathrm{e^+e^-}}\xspace}
\providecommand{\MM}{\ensuremath{\mu^+\mu^-}\xspace}
\providecommand{\TT}{\ensuremath{\tau^+\tau^-}\xspace}

\providecommand{\HGG}{\ensuremath{\mathrm{H}\to\gamma\gamma}}
\providecommand{\GAMJET}{\ensuremath{\gamma + \text{jet}}}
\providecommand{\PPTOJETS}{\ensuremath{\mathrm{pp}\to\text{jets}}}
\providecommand{\PPTOGG}{\ensuremath{\mathrm{pp}\to\gamma\gamma}}
\providecommand{\PPTOGAMJET}{\ensuremath{\mathrm{pp}\to\gamma + \mathrm{jet}}}
\providecommand{\MH}{\ensuremath{M_{\mathrm{H}}}}
\providecommand{\RNINE}{\ensuremath{R_\mathrm{9}}}
\providecommand{\DR}{\ensuremath{\Delta R}}

%

\providecommand{\ga}{\ensuremath{\gtrsim}}
\providecommand{\la}{\ensuremath{\lesssim}}
\providecommand{\swsq}{\ensuremath{\sin^2\theta_\cmsSymbolFace{W}}\xspace}
\providecommand{\cwsq}{\ensuremath{\cos^2\theta_\cmsSymbolFace{W}}\xspace}
\providecommand{\tanb}{\ensuremath{\tan\beta}\xspace}
\providecommand{\tanbsq}{\ensuremath{\tan^{2}\beta}\xspace}
\providecommand{\sidb}{\ensuremath{\sin 2\beta}\xspace}
\providecommand{\alpS}{\ensuremath{\alpha_S}\xspace}
\providecommand{\alpt}{\ensuremath{\tilde{\alpha}}\xspace}

\providecommand{\QL}{\ensuremath{\cmsSymbolFace{Q}_\cmsSymbolFace{L}}\xspace}
\providecommand{\sQ}{\ensuremath{\tilde{\cmsSymbolFace{Q}}}\xspace}
\providecommand{\sQL}{\ensuremath{\tilde{\cmsSymbolFace{Q}}_\cmsSymbolFace{L}}\xspace}
\providecommand{\ULC}{\ensuremath{\cmsSymbolFace{U}_\cmsSymbolFace{L}^\cmsSymbolFace{C}}\xspace}
\providecommand{\sUC}{\ensuremath{\tilde{\cmsSymbolFace{U}}^\cmsSymbolFace{C}}\xspace}
\providecommand{\sULC}{\ensuremath{\tilde{\cmsSymbolFace{U}}_\cmsSymbolFace{L}^\cmsSymbolFace{C}}\xspace}
\providecommand{\DLC}{\ensuremath{\cmsSymbolFace{D}_\cmsSymbolFace{L}^\cmsSymbolFace{C}}\xspace}
\providecommand{\sDC}{\ensuremath{\tilde{\cmsSymbolFace{D}}^\cmsSymbolFace{C}}\xspace}
\providecommand{\sDLC}{\ensuremath{\tilde{\cmsSymbolFace{D}}_\cmsSymbolFace{L}^\cmsSymbolFace{C}}\xspace}
\providecommand{\LL}{\ensuremath{\cmsSymbolFace{L}_\cmsSymbolFace{L}}\xspace}
\providecommand{\sL}{\ensuremath{\tilde{\cmsSymbolFace{L}}}\xspace}
\providecommand{\sLL}{\ensuremath{\tilde{\cmsSymbolFace{L}}_\cmsSymbolFace{L}}\xspace}
\providecommand{\ELC}{\ensuremath{\cmsSymbolFace{E}_\cmsSymbolFace{L}^\cmsSymbolFace{C}}\xspace}
\providecommand{\sEC}{\ensuremath{\tilde{\cmsSymbolFace{E}}^\cmsSymbolFace{C}}\xspace}
\providecommand{\sELC}{\ensuremath{\tilde{\cmsSymbolFace{E}}_\cmsSymbolFace{L}^\cmsSymbolFace{C}}\xspace}
\providecommand{\sEL}{\ensuremath{\tilde{\cmsSymbolFace{E}}_\cmsSymbolFace{L}}\xspace}
\providecommand{\sER}{\ensuremath{\tilde{\cmsSymbolFace{E}}_\cmsSymbolFace{R}}\xspace}
\providecommand{\sFer}{\ensuremath{\tilde{\cmsSymbolFace{f}}}\xspace}
\providecommand{\sQua}{\ensuremath{\tilde{\cmsSymbolFace{q}}}\xspace}
\providecommand{\sUp}{\ensuremath{\tilde{\cmsSymbolFace{u}}}\xspace}
\providecommand{\suL}{\ensuremath{\tilde{\cmsSymbolFace{u}}_\cmsSymbolFace{L}}\xspace}
\providecommand{\suR}{\ensuremath{\tilde{\cmsSymbolFace{u}}_\cmsSymbolFace{R}}\xspace}
\providecommand{\sDw}{\ensuremath{\tilde{\cmsSymbolFace{d}}}\xspace}
\providecommand{\sdL}{\ensuremath{\tilde{\cmsSymbolFace{d}}_\cmsSymbolFace{L}}\xspace}
\providecommand{\sdR}{\ensuremath{\tilde{\cmsSymbolFace{d}}_\cmsSymbolFace{R}}\xspace}
\providecommand{\sTop}{\ensuremath{\tilde{\cmsSymbolFace{t}}}\xspace}
\providecommand{\stL}{\ensuremath{\tilde{\cmsSymbolFace{t}}_\cmsSymbolFace{L}}\xspace}
\providecommand{\stR}{\ensuremath{\tilde{\cmsSymbolFace{t}}_\cmsSymbolFace{R}}\xspace}
\providecommand{\stone}{\ensuremath{\tilde{\cmsSymbolFace{t}}_1}\xspace}
\providecommand{\sttwo}{\ensuremath{\tilde{\cmsSymbolFace{t}}_2}\xspace}
\providecommand{\sBot}{\ensuremath{\tilde{\cmsSymbolFace{b}}}\xspace}
\providecommand{\sbL}{\ensuremath{\tilde{\cmsSymbolFace{b}}_\cmsSymbolFace{L}}\xspace}
\providecommand{\sbR}{\ensuremath{\tilde{\cmsSymbolFace{b}}_\cmsSymbolFace{R}}\xspace}
\providecommand{\sbone}{\ensuremath{\tilde{\cmsSymbolFace{b}}_1}\xspace}
\providecommand{\sbtwo}{\ensuremath{\tilde{\cmsSymbolFace{b}}_2}\xspace}
\providecommand{\sLep}{\ensuremath{\tilde{\cmsSymbolFace{l}}}\xspace}
\providecommand{\sLepC}{\ensuremath{\tilde{\cmsSymbolFace{l}}^\cmsSymbolFace{C}}\xspace}
\providecommand{\sEl}{\ensuremath{\tilde{\cmsSymbolFace{e}}}\xspace}
\providecommand{\sElC}{\ensuremath{\tilde{\cmsSymbolFace{e}}^\cmsSymbolFace{C}}\xspace}
\providecommand{\seL}{\ensuremath{\tilde{\cmsSymbolFace{e}}_\cmsSymbolFace{L}}\xspace}
\providecommand{\seR}{\ensuremath{\tilde{\cmsSymbolFace{e}}_\cmsSymbolFace{R}}\xspace}
\providecommand{\snL}{\ensuremath{\tilde{\nu}_L}\xspace}
\providecommand{\sMu}{\ensuremath{\tilde{\mu}}\xspace}
\providecommand{\sNu}{\ensuremath{\tilde{\nu}}\xspace}
\providecommand{\sTau}{\ensuremath{\tilde{\tau}}\xspace}
\providecommand{\Glu}{\ensuremath{\cmsSymbolFace{g}}\xspace}
\providecommand{\sGlu}{\ensuremath{\tilde{\cmsSymbolFace{g}}}\xspace}
\providecommand{\Wpm}{\ensuremath{\cmsSymbolFace{W}^{\pm}}\xspace}
\providecommand{\sWpm}{\ensuremath{\tilde{\cmsSymbolFace{W}}^{\pm}}\xspace}
\providecommand{\Wz}{\ensuremath{\cmsSymbolFace{W}^{0}}\xspace}
\providecommand{\sWz}{\ensuremath{\tilde{\cmsSymbolFace{W}}^{0}}\xspace}
\providecommand{\sWino}{\ensuremath{\tilde{\cmsSymbolFace{W}}}\xspace}
\providecommand{\Bz}{\ensuremath{\cmsSymbolFace{B}^{0}}\xspace}
\providecommand{\sBz}{\ensuremath{\tilde{\cmsSymbolFace{B}}^{0}}\xspace}
\providecommand{\sBino}{\ensuremath{\tilde{\cmsSymbolFace{B}}}\xspace}
\providecommand{\Zz}{\ensuremath{\cmsSymbolFace{Z}^{0}}\xspace}
\providecommand{\sZino}{\ensuremath{\tilde{\cmsSymbolFace{Z}}^{0}}\xspace}
\providecommand{\sGam}{\ensuremath{\tilde{\gamma}}\xspace}
\providecommand{\chiz}{\ensuremath{\tilde{\chi}^{0}}\xspace}
\providecommand{\chip}{\ensuremath{\tilde{\chi}^{+}}\xspace}
\providecommand{\chim}{\ensuremath{\tilde{\chi}^{-}}\xspace}
\providecommand{\chipm}{\ensuremath{\tilde{\chi}^{\pm}}\xspace}
\providecommand{\Hone}{\ensuremath{\cmsSymbolFace{H}_\cmsSymbolFace{d}}\xspace}
\providecommand{\sHone}{\ensuremath{\tilde{\cmsSymbolFace{H}}_\cmsSymbolFace{d}}\xspace}
\providecommand{\Htwo}{\ensuremath{\cmsSymbolFace{H}_\cmsSymbolFace{u}}\xspace}
\providecommand{\sHtwo}{\ensuremath{\tilde{\cmsSymbolFace{H}}_\cmsSymbolFace{u}}\xspace}
\providecommand{\sHig}{\ensuremath{\tilde{\cmsSymbolFace{H}}}\xspace}
\providecommand{\sHa}{\ensuremath{\tilde{\cmsSymbolFace{H}}_\cmsSymbolFace{a}}\xspace}
\providecommand{\sHb}{\ensuremath{\tilde{\cmsSymbolFace{H}}_\cmsSymbolFace{b}}\xspace}
\providecommand{\sHpm}{\ensuremath{\tilde{\cmsSymbolFace{H}}^{\pm}}\xspace}
\providecommand{\hz}{\ensuremath{\cmsSymbolFace{h}^{0}}\xspace}
\providecommand{\Hz}{\ensuremath{\cmsSymbolFace{H}^{0}}\xspace}
\providecommand{\Az}{\ensuremath{\cmsSymbolFace{A}^{0}}\xspace}
\providecommand{\Hpm}{\ensuremath{\cmsSymbolFace{H}^{\pm}}\xspace}
\providecommand{\sGra}{\ensuremath{\tilde{\cmsSymbolFace{G}}}\xspace}
\providecommand{\mtil}{\ensuremath{\tilde{m}}\xspace}
\providecommand{\rpv}{\ensuremath{\rlap{\kern.2em/}R}\xspace}
\providecommand{\LLE}{\ensuremath{LL\bar{E}}\xspace}
\providecommand{\LQD}{\ensuremath{LQ\bar{D}}\xspace}
\providecommand{\UDD}{\ensuremath{\overline{UDD}}\xspace}
\providecommand{\Lam}{\ensuremath{\lambda}\xspace}
\providecommand{\Lamp}{\ensuremath{\lambda'}\xspace}
\providecommand{\Lampp}{\ensuremath{\lambda''}\xspace}
\providecommand{\spinbd}[2]{\ensuremath{\bar{#1}_{\dot{#2}}}\xspace}

\providecommand{\MD}{\ensuremath{{M_\mathrm{D}}}\xspace}
\providecommand{\Mpl}{\ensuremath{{M_\mathrm{Pl}}}\xspace}
\providecommand{\Rinv} {\ensuremath{{R}^{-1}}\xspace} 
\newcommand {\roots}    {\ensuremath{\sqrt{s}}}
\newcommand {\rootsNN}  {\ensuremath{\sqrt{s_{_{NN}}}}}
\newcommand {\dndy}     {\ensuremath{dN/dy}}
\newcommand {\dnchdy}   {\ensuremath{dN_{\mathrm{ch}}/dy}}
\newcommand {\dndeta}   {\ensuremath{dN/d\eta}}
\newcommand {\dnchdeta} {\ensuremath{dN_{\mathrm{ch}}/d\eta}}
\newcommand {\dndpt}    {\ensuremath{dN/d\pt}}
\newcommand {\dnchdpt}  {\ensuremath{dN_{\mathrm{ch}}/d\pt}}
\newcommand {\deta}     {\ensuremath{\Delta\eta}}
\newcommand {\dphi}     {\ensuremath{\Delta\phi}}
\newcommand {\AJ}       {\ensuremath{A_J}}
\newcommand {\npart}    {\ensuremath{N_{\rm part}}}
\newcommand {\ncoll}    {\ensuremath{N_{\rm coll}}}

\newcommand {\pp}    {\mbox{pp}}
\newcommand {\ppbar} {\mbox{p\={p}}}
\newcommand {\pbarp} {\mbox{p\={p}}}
\newcommand {\PbPb}  {\mbox{PbPb}}

\newcommand{\m}{\ensuremath{\,\text{m}}\xspace}

\newcommand {\naive}    {na\"{\i}ve}
\providecommand{\PHOJET} {\textsc{phojet}\xspace}

\def\d{\mathrm{d}}

\providecommand{\PKzS}{\ensuremath{\mathrm{K^0_S}}}
\providecommand{\Pp}{\ensuremath{\mathrm{p}}}
\providecommand{\Pap}{\ensuremath{\mathrm{\overline{p}}}}
\providecommand{\PgL}{\ensuremath{\Lambda}}
\providecommand{\PagL}{\ensuremath{\overline{\Lambda}}}
\providecommand{\PgS}{\ensuremath{\Sigma}}
\providecommand{\PgSm}{\ensuremath{\Sigma^-}}
\providecommand{\PgSp}{\ensuremath{\Sigma^+}}
\providecommand{\PagSm}{\ensuremath{\overline{\Sigma}^-}}
\providecommand{\PagSp}{\ensuremath{\overline{\Sigma}^+}}

\hyphenation{off-line}

\cmsNoteHeader{HIN-10-004} 
\title{Observation and studies of jet quenching in PbPb collisions
at $\sqrt{s_{_{NN}}}=2.76$~TeV}

\address[cern]{CERN}
\author[cern]{The CMS Collaboration}

\date{\today}

\abstract{Jet production in PbPb collisions at a nucleon-nucleon center-of-mass
energy of 2.76~TeV was studied with the CMS detector at the LHC,
using a data sample corresponding to an integrated luminosity of $6.7\,\mu$b$^{-1}$.
Jets are reconstructed using the energy deposited in the CMS calorimeters and
studied
as a function of collision centrality.
With increasing collision centrality, a striking imbalance in dijet transverse momentum is
observed, consistent with jet quenching.
The observed effect extends from the lower cut-off used in this study
(jet $\pt = 120$~\GeVc) up to the statistical limit of the available
data sample (jet $\pt \approx 210$~\GeVc).
Correlations of charged particle tracks with jets indicate that the momentum imbalance
is accompanied by a softening of the fragmentation pattern of the
second most energetic, away-side jet.
The dijet momentum balance is recovered when integrating
low transverse momentum particles distributed over a wide angular range relative
to the direction of the away-side jet.
}

\hypersetup{%
pdfauthor={CMS Collaboration},%
pdftitle={Observation and studies of jet quenching in PbPb collisions
at nucleon-nucleon center-of-mass energy = 2.76 TeV},%
pdfsubject={CMS},%
pdfkeywords={CMS, physics, heavy ion, jet}}

\maketitle 

\section{Introduction}
\label{sec:introduction}

High-energy collisions of heavy ions allow  the fundamental theory of the strong interaction --- Quantum Chromodynamics (QCD) ---  
to be studied under extreme temperature and density conditions. A new form of matter 
\cite{Shuryak:1977ut,Collins:1974ky,Cabibbo:1975ig,Freedman:1976ub} formed at energy densities above 
$\sim$1~GeV/fm$^3$  is predicted in Lattice QCD calculations \cite{Karsch:2003jg}. This quark-gluon plasma (QGP) 
consists of an extended volume of deconfined and chirally-symmetric quarks and gluons.

Heavy ion collisions at the Large Hadron Collider (LHC) are expected to produce matter at 
energy densities exceeding any previously explored in experiments conducted at particle accelerators. 
One of the first experimental signatures suggested for QGP studies
was the suppression of high-transverse-momentum (\pt) hadron yields resulting from energy loss suffered
by hard-scattered partons passing through the medium~\cite{Bjorken:1982tu}. 
This parton energy loss is often referred to as ``jet quenching''.
The energy lost by a parton provides fundamental information on the thermodynamical
and transport properties of the traversed medium, which is now
believed to be strongly coupled as opposed to an ideal gas of quarks and gluons 
(recent reviews: ~\cite{CasalderreySolana:2007zz,d'Enterria:2009am}).
Results from nucleus-nucleus collisions at the Relativistic Heavy Ion Collider 
(RHIC)~\cite{Adcox:2004mh,Adams:2005dq,Back:2004je,Arsene:2004fa} 
have shown evidence for the quenching effect through the suppression of
inclusive high-$\pt$ hadron production and the modification of  high-$\pt$ dihadron 
angular correlations when compared to the corresponding results in much smaller systems, especially proton-proton collisions.
Preliminary results for fully reconstructed jets at RHIC, measured in AuAu collisions at $\sqrt{s_{_{NN}}}=200$~GeV~\cite{Salur:2008hs, Putschke:2008wn,Bruna:2009em, Ploskon:2009zd}, also hint at  broadened jet shapes due to medium-induced gluon radiation.

Studying the modification of jets has long been proposed as a particularly useful tool for probing the
QGP properties~\cite{Appel:1985dq,Blaizot:1986ma}. 
Of particular interest are the dominant ``dijets'', consisting of the most energetic (``leading'')
and second most energetic (``subleading'') jets.
At leading order (LO) and in the absence of parton energy loss,
the two jets have equal $\pt$ with respect to the beam axis 
and are emitted very close to back-to-back in azimuth 
($\Delta \varphi_\text{dijet} = \left |\varphi_\text{jet1} - \varphi_\text{jet2} \right |\approx \pi$). 
However, medium-induced gluon emission in the final state can significantly 
alter the energy balance between the two highest-$\pt$ jets
and may give rise to large deviations from $\Delta \varphi_\text{dijet} \approx \pi$.
Such medium effects in nuclear interactions are expected to be much larger than those due to higher-order
gluon radiation, which is also present for jet events in pp collisions.
The study of medium-induced modifications of dijet properties can
therefore shed light on the transport properties of the QCD medium formed in heavy ion
collisions.

The dijet analysis presented in this paper was performed using the data collected in 2010 from  PbPb collisions at a nucleon-nucleon center-of-mass energy of $\sqrt{s_{_{NN}}}=2.76$~TeV at the Compact Muon Solenoid (CMS) detector.  The CMS detector has 
a solid angle acceptance of nearly 4$\pi$ and is designed to measure jets and energy flow, an ideal feature for studying heavy ion collisions. A total integrated (PbPb) 
luminosity of 8.7~$\mu$b$^{-1}$ was collected, of which 6.7~$\mu$b$^{-1}$ has been included
in this analysis. Recently, related results on a smaller data sample (1.7~$\mu$b$^{-1}$) 
have been reported by ATLAS~\cite{Collaboration:2010bu}.

\begin{figure}[hbtp]
  \begin{center}
    \includegraphics[trim=2.5cm 0cm 0cm 0cm, clip=true, width=0.8\textwidth]{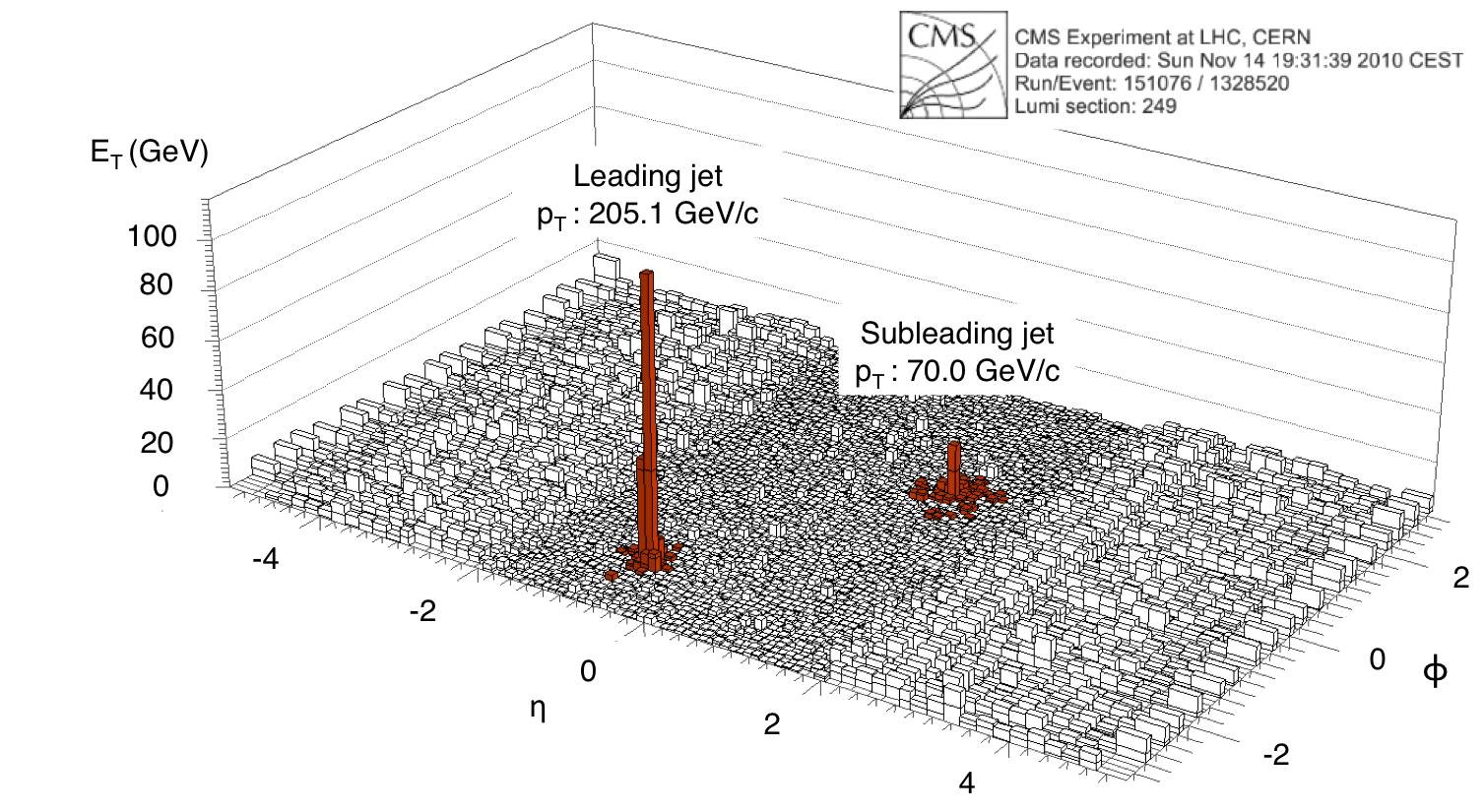}
    \caption{Example of an unbalanced dijet in a PbPb collision event at $\sqrt{s_{_{NN}}}=2.76$~TeV. 
Plotted is the summed transverse energy in the electromagnetic and hadron calorimeters vs. $\eta$ and $\phi$, 
with the identified jets highlighted in red, and labeled with the corrected jet transverse momentum.}
    \label{fig:eventDisplay}
  \end{center}
\end{figure}

Jets were reconstructed based on their energy deposits in the CMS calorimeters.
In general, the jet quenching effect on partons traversing the medium
with different path lengths will lead to modifications in the observed dijet energy balance
due to a combination of two effects: the radiated energy can fall outside the area used 
for the determination of the jet energy, and the energy can be shifted towards low
momentum particles, which will not be detected in the calorimetric energy measurement. 
Such unbalanced events are easy to detect visually even at the level of event displays, and numerous examples 
were in fact seen during the first days of data taking (e.g. Fig.~\ref{fig:eventDisplay}).  

The data provide information on the evolution of the dijet imbalance as a
function of both collision centrality (i.e., the degree of overlap of the two
colliding nuclei) and the energy of the leading jet. By 
correlating the dijets detected in the calorimeters with charged hadrons 
reconstructed in the high-resolution tracker system, the modification of the jet 
fragmentation pattern can be studied in detail, thus providing a deeper 
insight into the dynamics of the jet quenching phenomenon.

The paper is organized as follows: the experimental setup, event triggering, selection and characterization, and jet reconstruction are described in Section~\ref{sec:experimental_method}. Section~\ref{sec:results} 
presents the results and a discussion of systematic uncertainties, followed by a summary in Section~\ref{sec:summary}.

\section{Experimental method}
\label{sec:experimental_method}

The CMS detector is described in detail
elsewhere~\cite{bib_CMS}.  The calorimeters provide hermetic coverage over a large
range of pseudorapidity, $|\eta| < 5.2$, where $\eta = -\text{ln [
tan}(\theta/2)]$ and $\theta$ is the polar angle relative to the
particle beam. 
In this study, jets are identified primarily using
the energy deposited in the lead-tungstate crystal electromagnetic
calorimeter (ECAL) and the brass/scintillator hadron calorimeter (HCAL)
covering $|\eta| < 3$. 
In addition, a steel/quartz-fiber Cherenkov
calorimeter, called Hadron Forward (HF), covers the forward rapidities 
$3<|\eta|<5.2$ and  is  used to determine the centrality of the PbPb collision.
Calorimeter cells are grouped in projective
towers of granularity in pseudorapidity and azimuthal angle given by 
$\Delta \eta \times \Delta \varphi =0.087\times0.087$ at central rapidities, having a coarser segmentation at forward
rapidities.  The central calorimeters are embedded in a solenoid with 3.8~T central magnetic field. The event display
shown in Fig.~\ref{fig:eventDisplay} illustrates the projective calorimeter tower granularity over the full pseudorapidity range.
The CMS tracking system, located inside the calorimeter, consists of pixel and silicon-strip layers covering  $|\eta| < 
2.5$, and provides track reconstruction down to $\pt \approx 100$~\MeVc, with
a track momentum resolution of about 1\% at $\pt = 100$~\GeVc. A set of scintillator tiles, the Beam Scintillator Counters (BSC), 
are mounted on the inner side of the HF calorimeters for triggering and beam-halo rejection.
CMS uses a right-handed coordinate system, with the
origin located at the nominal collision point at the center of the detector, the $x$-axis
pointing towards the center of the LHC ring, the $y$-axis pointing up
(perpendicular to the LHC plane), and the $z$-axis along the
counterclockwise beam direction.
The detailed Monte Carlo (MC) simulation of the CMS detector response is based on {\GEANTfour} 
\cite{bib_geant}.

\subsection{Data samples and triggers}

The expected cross section for hadronic inelastic PbPb collisions at $\sqrt{s_{_{NN}}}=2.76$~TeV
is 7.65~b, corresponding to the chosen Glauber MC parameters described in Section~\ref{sec:centrality}. 
In addition, there is a sizable contribution from large impact parameter ultra-peripheral collisions (UPC)
that  lead to the electromagnetic breakup of one, or both, of the Pb nuclei~\cite{Djuvsland:2010qs}.
particles per unit of pseudorapidity, depending on the impact parameter.

\begin{figure}[hbtp]
\begin{center}
\includegraphics[width=0.45\textwidth, viewport=0 -3 567 490]{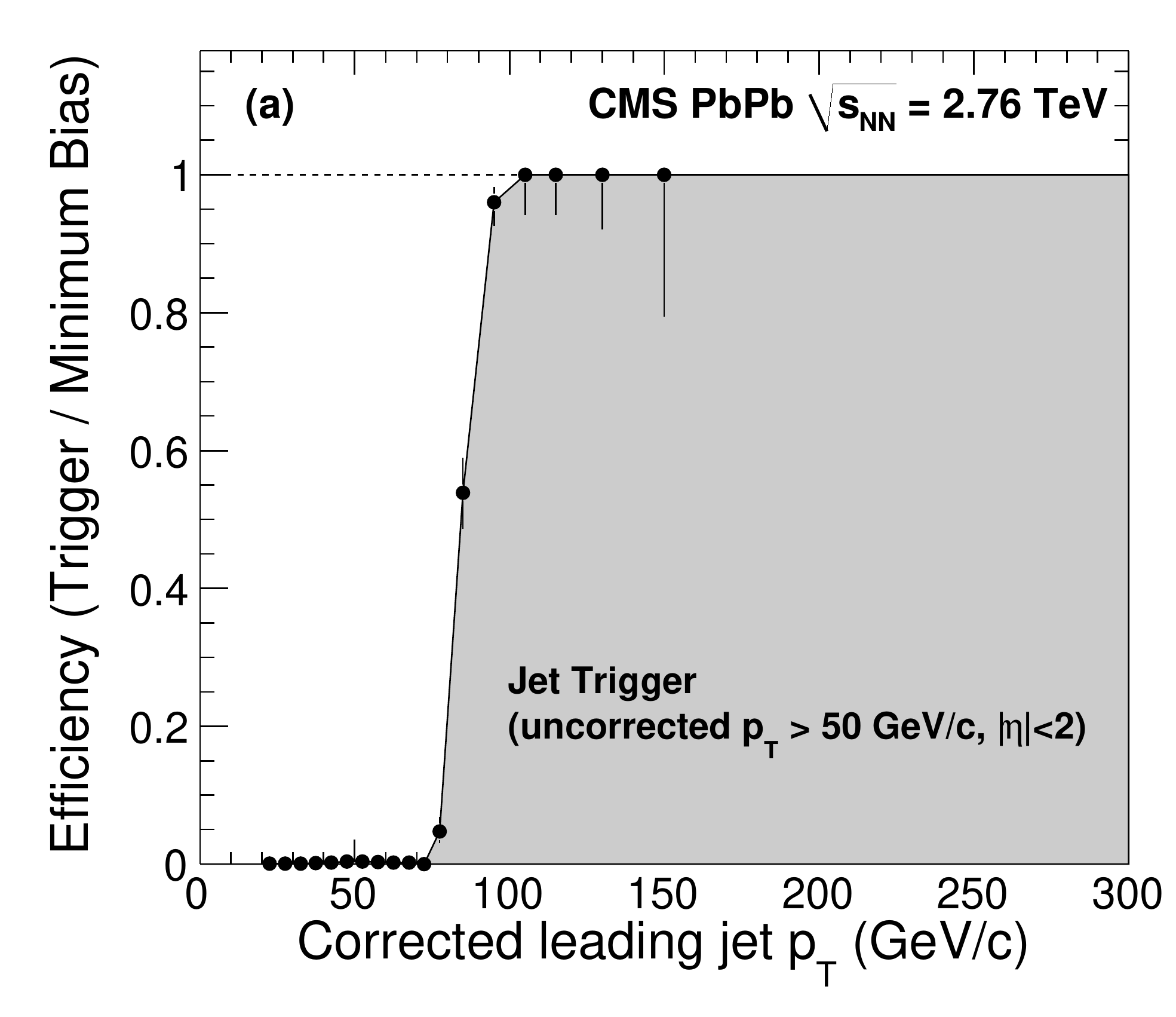}
\includegraphics[width=0.49\textwidth]{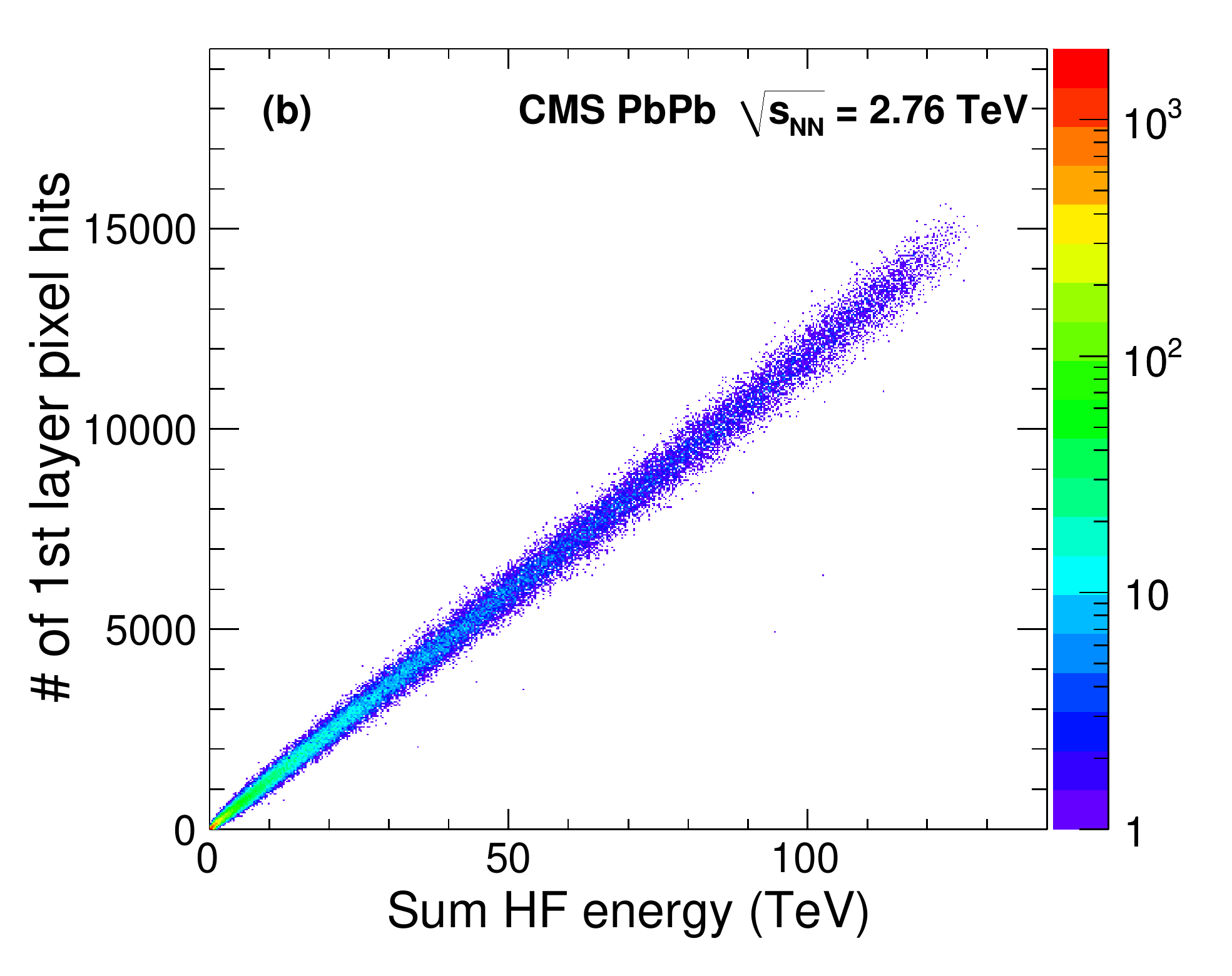}
\caption{(a) Efficiency curve for the HLT 50~\GeVc single-jet trigger, as a function of the
corrected leading jet transverse momentum. Error bars shown are statistical. (b) Correlation between the number of pixel hits and HF total energy for a single run containing 60\,k minimum bias events, after selections as described in the text.
}
\label{pixelhits}
\label{turnon}
\end{center}
\end{figure}

For online event selection, 
CMS uses a two-level trigger system:
Level-1 (L1) and High Level Trigger (HLT). The events for this analysis were
selected using an inclusive single-jet trigger that required an
L1 jet with  $\pt > 30$~\GeVc  and an HLT jet 
with  $\pt > 50$~\GeVc , where neither $\pt$ value was
corrected for the $\pt$-dependent calorimeter energy response 
discussed in Section~\ref{sec:jet_performance}.
The efficiency of the jet trigger 
is shown in Fig.~\ref{turnon}\,(a) for leading jets with $|\eta| <2 $ 
as a function of their corrected $\pt$. 
The efficiency is defined as the ratio of the number 
of triggered events over the number of minimum bias events (described below).
The trigger becomes fully efficient for collisions with a leading jet
with corrected $\pt$ greater than 100~\GeVc.

In addition to the jet data sample, a minimum bias event sample was collected using  
coincidences between the trigger signals from the $+z$ and $-z$ sides of either the BSC 
or the HF.
This trigger has an efficiency of more than 97\%  for hadronic inelastic collisions.
In order to suppress non-collision related noise, cosmic rays, 
double-firing triggers, and beam backgrounds, the minimum bias and jet triggers used in this analysis
were required to fire in time with the presence of both colliding ion bunches in the interaction region.
It was checked that the events selected by the jet trigger described above also satisfy all triggers and selections 
imposed for minimum bias events.  The total hadronic collision rate varied between 1 and 210 Hz, 
depending on the number of colliding bunches (between 1$\times$1 and 129$\times$129) and on the bunch intensity.

\subsection{Event selection}
\label{sec:event_selection}

In order to select a pure sample of inelastic hadronic collisions for analysis, 
a number of 
offline selections were applied to the triggered event sample,
removing contaminations from UPC events and non-collision beam backgrounds (e.g. beam-gas).
Table~\ref{evselcuts} shows the number of events remaining after the various selection criteria are applied.
First, beam halo events were vetoed based on the timing of the $+z$ and $-z$ BSC signals.
Then, to veto UPC and beam-gas events, an offline HF coincidence of at least three towers on each side of the 
interaction point was required, with a total deposited energy of at least 3~GeV.
Next, a reconstructed vertex was required with at least two tracks of $\pt > 75$~\MeVc, 
consistent with the transverse beam spot position and the expected collision region along the $z$-axis.
Finally, to further reject beam-gas and beam-scraping events, the length of pixel clusters along the beam direction 
were required to be compatible with particles originating from the primary vertex.  This last selection is identical to the one used for the study of charged hadron pseudorapidity density and
\pt\  spectrum in 7~TeV \pp\ collisions ~\cite{Khachatryan:2010us}.
Figure~\ref{pixelhits}\,(b) shows the correlation between the total energy deposited in the HF calorimeters
and the number of hits in the first layer of the silicon pixel barrel detector after these event selections. 
A tight correlation between the two detectors is observed, 
with very few of the events showing HF energy deposits that deviate significantly
(at any given number of pixel hits) from the expectations for hadronic PbPb collisions.
This correlation is important to verify the selection of a pure collision event sample, 
and also to validate the HF energy sum as a measure of event centrality (Section \ref{sec:centrality}).

Starting from inelastic hadron collisions based on the selections described above, 
the basic offline selection of events for the analysis is the presence of a leading calorimeter jet
in the pseudorapidity range of $|\eta| <  2$ with a corrected jet $\pt > 120$~\GeVc. 
By selecting these leading jets we avoid possible biases due to inefficiencies
close to the trigger threshold. Furthermore, the selection of a rather large leading jet momentum expands
the range of jet momentum imbalances that can be observed between the leading and subleading jets,
as the subleading jets need a minimum momentum of $\pt > $ 35--50~\GeVc to be reliably detected above the
high-multiplicity underlying event in PbPb collisions (Section~\ref{sec:jet_reconstruction}).
In order to ensure high quality dijet selection,  kinematic selection cuts were applied. The azimuthal angle between the leading and subleading jet was required to be at least $2\pi/3$. Also, we require a minimum \pt\ of $p_{\mathrm{T},1} > 120$~\GeVc for leading jets and of $p_{\mathrm{T},2} > 50$~\GeVc  for subleading jets. 
No explicit requirement is made either on the presence or absence of a third jet in the event.
Prior to jet finding on the selected events, a small contamination of
noise events from uncharacteristic ECAL and HCAL detector responses was removed using signal
timing, energy distribution, and pulse-shape information \cite{ref:EGM-10-002,Chatrchyan:2009hy}. As a result,
about 2.4\% of the events were removed from the sample.  

\begin{table}[]
\begin{center}
\caption{Event selection criteria used for this analysis.  The percentage of events remaining after
each criterion, listed in the last column, are with respect to the previous criterion (the event selection criteria are applied in the indicated sequence). 
\protect\\ }
\label{evselcuts} 
\begin{tabular}{|l|r|r|}
\hline
Criterion & Events remaining & \% of events remaining \\
\hline
Jet triggered events ($\pt^{\mathrm{uncorr}}>50$~GeV/$c$) & 149\,k        & 100.00\\
No beam halo, based on the BSC  & 148\,k        & 99.61\\
HF offline coincidence          & 111\,k        & 74.98\\
Reconstructed vertex            & 110\,k        & 98.97\\
Beam-gas removal                & 110\,k        & 99.78\\
ECAL cleaning                   & 107\,k        & 97.66\\
HCAL cleaning                   & 107\,k        & 99.97\\
$\geq2$ jets with $\pt>35$~\GeVc and $|\eta|<2$ & 71.9k & 67.07 \\
Leading jet $p_{\mathrm{T},1}>120$~\GeVc        & 4\,216  & 5.86 \\
Subleading jet $p_{\mathrm{T},2}>50$~\GeVc      & 3\,684  & 87.38 \\
$\Delta\phi_{12}$ of 2 jets $>2\pi/3$ & 3\,514     & 95.39 \\
\hline
\end{tabular}
\end{center}
\end{table}

\subsection{Centrality determination}
\label{sec:centrality}

\begin{figure}[ht]
\begin {center}
\mbox{
\includegraphics[width=0.48\linewidth]{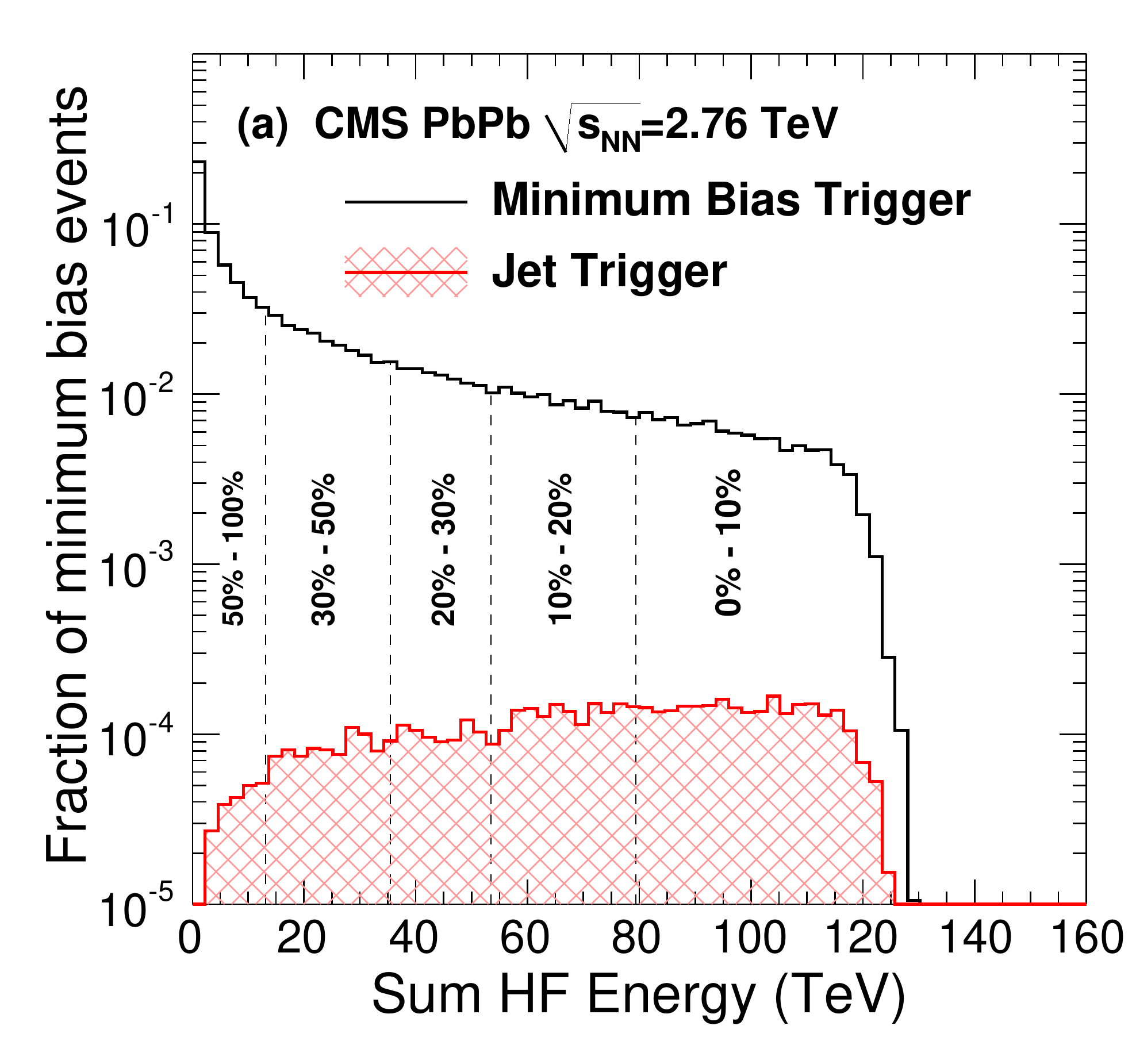}}\hfill
\mbox{\includegraphics[width=0.48\linewidth]{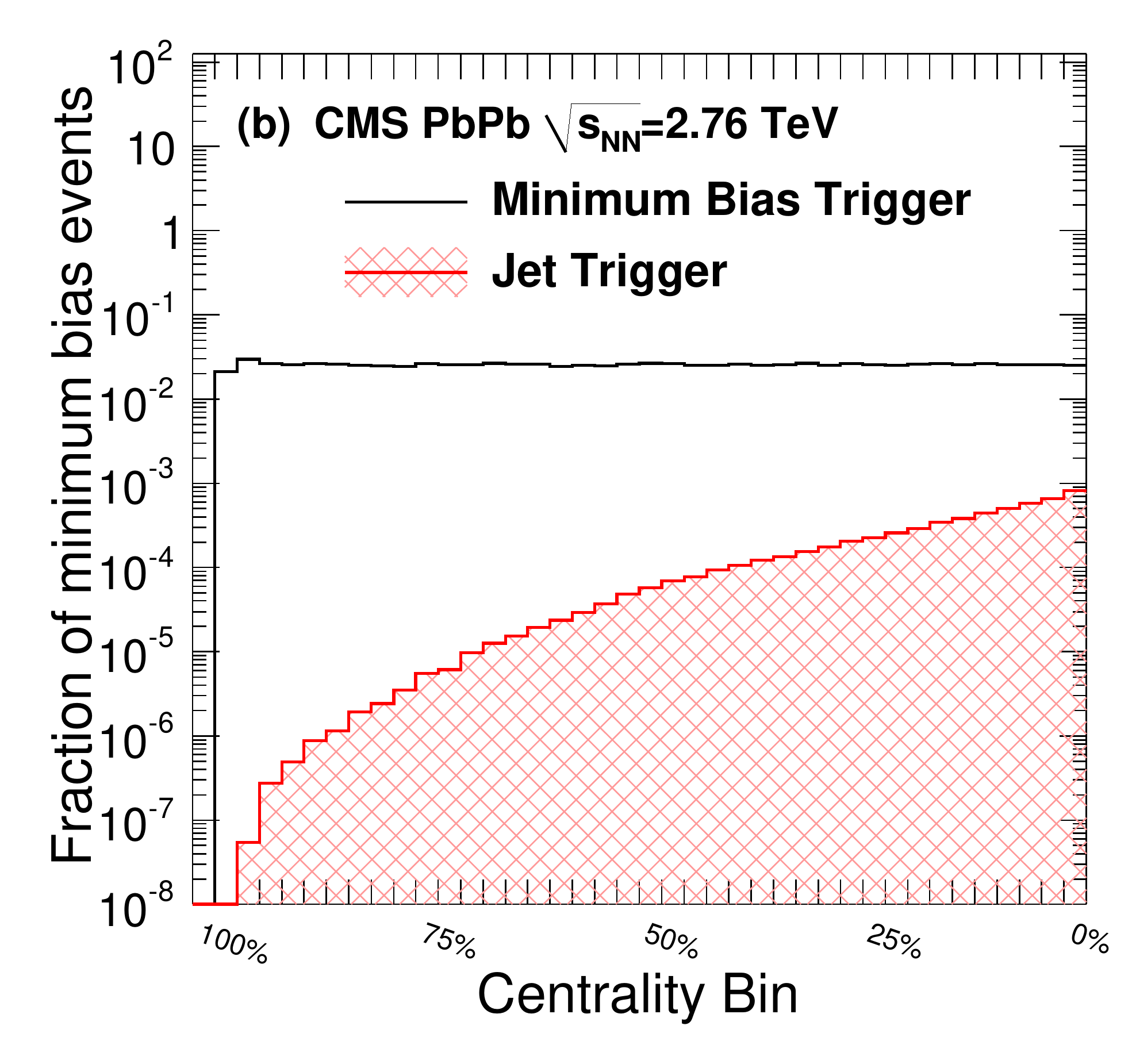}}
\caption{
(a) Probability distribution of the total HF energy for minimum bias collisions
(black open histogram). The five regions correspond to the centrality ranges used
in this analysis. Also shown is the HF energy distribution for the subset of events passing 
the HLT jet trigger (red hatched histogram). (b)
Distribution of the fraction of events in the 40 centrality bins for minimum bias
(black open histogram) and HLT jet triggered (red hatched histogram)
events. The centrality-bin labels run from 100\% for the most
peripheral to 0\% for the most central events.}
\label{fig:HF_cent}
\end{center}
\end{figure}

For the analysis of PbPb events, it is important to know the ``centrality'' of the collision,
i.e., whether the overlap of the two colliding nuclei is large or small. 
In this analysis, the observable used to determine centrality is the total energy 
from both  HF calorimeters.  
The distribution of the HF signal used
in this analysis is shown in Fig.~\ref{fig:HF_cent}~(a). The shape 
of the energy distribution is characteristic of all observables
related to (soft) particle production in heavy ion collisions. The more frequent
peripheral events with large impact parameter produce very few particles, 
while the central ones with small impact parameter produce many more 
particles because of the increased number of nucleon-nucleon interactions.

The distribution of this total energy was used to divide the event sample 
into 40 centrality bins, each representing 2.5\% of the total nucleus-nucleus interaction 
cross section. Because of inefficiencies in the minimum bias trigger and event selection, 
the measured multiplicity distribution does not represent the full interaction cross 
section. MC simulations were used to estimate the distribution in the regions where 
events are lost. Comparing the simulated distribution to the measured distribution,
it is estimated that the minimum bias trigger and event 
selection efficiency is $97 \pm 3$\%.
 
For the jet analysis, these
fine-grained bins were combined into five larger bins corresponding to the 
most central 10\% of the events (i.e., smallest impact parameter), the next most
central 10\% of the events (denoted 10--20\%), and further bins 
corresponding to the 20--30\%, 30--50\%, and 50--100\% selections of the total 
hadronic cross section.

Simulations can be used to correlate centrality, as quantified using the fraction of the total interaction
cross section, with more detailed properties of the collision. The two
most commonly used physical quantities are the total number of nucleons in the two
lead ($^{208}$Pb) nuclei which experienced at least one inelastic collision, denoted \npart, and
the total number of binary nucleon-nucleon collisions, \ncoll. 

The centrality bins can be correlated to the impact parameter, $b$, and to
average values and variances of \npart\ and \ncoll\ 
using a calculation based on a Glauber model in which the nucleons are assumed to
follow straight-line trajectories as the nuclei collide (for a review, see \cite{Miller:2007ri}). The bin-to-bin smearing
of the results of these calculations due to the finite resolution and
fluctuations in the HF energy measurement was obtained from fully simulated
and reconstructed MC events generated with the {\sc{ampt}} event generator~\cite{Lin:2004en}.
Standard parameters of the Woods-Saxon function used to model the
distribution of nucleons in the Pb nuclei were used~\cite{DeJager:1987qc}.
The nucleon-nucleon inelastic cross section, which is used to determine how
close the nucleon trajectories need to be in order for an interaction to
occur, was taken to be $64\pm5$~mb, based on a fit of the existing data for total and
elastic cross-sections in proton-proton and proton-antiproton collisions~\cite{PDBook}. 
The uncertainties in the parameters involved in these calculations contribute to
the systematic uncertainty in \npart\ and \ncoll\ for a given bin. The other source of uncertainty in
the centrality parameters comes from the determination of the event selection efficiency.

Using the procedure outlined above, the mean and spread (RMS) values of the impact parameter, \npart, and \ncoll\ for the five bins used in this
analysis, and their systematic uncertainties, were extracted and are listed in Table ~\ref{tbl:VariablesForJetAnalysis}.
The RMS values for the centrality parameters are due to their correlation with the percentage cross section and the width of the chosen centrality bins.

It is important to note that the selection of rare processes, such as the production
of high-\pt\ jets, leads to a strong bias in the centrality distribution of the 
underlying events
towards more central collisions, for which \ncoll\ is very large. This can be seen in Fig.~\ref{fig:HF_cent}~(a), where the 
HF energy distribution for events selected by the jet trigger is shown in comparison
to that for minimum bias events. The bias can be seen more clearly in Fig.~\ref{fig:HF_cent}~(b),
where the distribution of minimum bias and jet-triggered events in the 40 centrality bins
is shown.

\begin{table}[htb]
\begin{center}
\caption{Mean and RMS values for the distributions of impact parameter, $b$, number of participating nucleons, \npart, 
and number of nucleon-nucleon collisions, \ncoll, for the centrality bins used in this analysis. The RMS values 
represent the spread of each quantity within the given bins due to the range of percentage cross section included. 
\label{tbl:VariablesForJetAnalysis} \protect\\ }
\begin{tabular}{|c|c|c|c|c|c|c|c|c|}
\hline
Centrality& 
$b$ mean (fm)&
$b$ RMS (fm)&
\npart\ mean &
\npart\ RMS &
\ncoll\ mean &
\ncoll\ RMS \\
\hline
0--10\% & 3.4 $\pm$ 0.1 & 1.2 & 355 $\pm$ 3 & 33 &1484 $\pm$ 120 & 241 \\
\hline
10--20\% & 6.0 $\pm$ 0.2 & 0.8 & 261 $\pm$ 4 &  30 &927 $\pm$ 82 & 183 \\
\hline
20--30\% & 7.8 $\pm$ 0.2 & 0.6  & 187 $\pm$ 5 & 23 &562 $\pm$ 53 & 124 \\
\hline
30--50\% & 9.9 $\pm$ 0.3 & 0.8 & 108 $\pm$ 5 & 27 &251 $\pm$ 28 & 101 \\
\hline
50--100\% & 13.6 $\pm$ 0.4  & 1.6 &  22 $\pm$ 2 & 19 &30 $\pm$ 5 & 35 \\
\hline
\end{tabular}
\end{center}
\end{table}

\subsection{Jet reconstruction in PbPb collisions}
\label{sec:jet_reconstruction}

\subsubsection{Jet algorithm}
The baseline jet reconstruction for heavy ion collisions in CMS is performed
with an iterative cone algorithm modified to subtract the soft underlying event  on an
event-by-event basis \cite{Kodolova:2007hd}. Each cone is selected with a radius 
$\Delta R=\sqrt{\Delta\phi^2+\Delta \eta^2} = 0.5$ around a seed of minimum transverse energy of 1 GeV.  The underlying event subtraction algorithm is a variant of an
iterative "noise/pedestal subtraction" technique~\cite{Ball:2007zza}. Initially, the mean value, $\langle E_{\rm cell} \rangle$, and
dispersion, $\sigma(E_{\rm cell})$, of the energies recorded in the calorimeter cells are calculated for
all rings of cells that have at least 0.3 GeV transverse energy deposit at constant pseudorapidity.
The algorithm subtracts $\langle E_{\rm cell}\rangle+\sigma(E_{\rm cell})$ from each cell.
If a cell energy is  negative after subtraction, the value is set to zero.
Subtracting the mean plus the dispersion, as opposed to simply the mean, 
compensates for the bias caused by the ``zeroing'' of negative-energy cells. 
Jets are then reconstructed, using a standard
iterative cone algorithm~\cite{Bayatian:2006zz,Huth:1990mi}, from the remaining cells with non-zero energy.
In a second
iteration, the pedestal function is recalculated using only calorimeter cells
outside the area covered by reconstructed high-$\pt$ jets ($\pt > 10$~\GeVc).
The threshold of 10~\GeVc was chosen in studies optimizing the final extracted jet \pt\ resolution.
The cell energies are updated with the new pedestal function (again subtracting mean plus dispersion)
and the jets are
reconstructed again, using the updated calorimeter cells. The performance of
this algorithm is documented in Ref. \cite{Kodolova:2007hd}.
Jet corrections for the calorimeter response have been applied, as determined in studies for \pp\
collisions~\cite{CMS-PAS-JME-10-010}. When applying the algorithm to \PbPb\ data, the subtracted background
energy for $R = 0.5$ jet cones 
ranges from 6--13~GeV for peripheral events (centrality bins 50--100\%) to 90--130~GeV for central collisions (0--10\%),
before applying jet energy scale corrections.

To perform a cross-check of the main results, the  anti-$k_{\rm T}$ algorithm~\cite{bib_antikt} with a 
resolution parameter of 0.5 was used to reconstruct jets, as was done for 
the \pp\ reference measurements presented here. The energy 
attributed to the underlying event was estimated and subtracted using the ``average energy per jet
area'' procedure provided by the {\sc fastjet} package~\cite{Cacciari:2008gn, Cacciari:2007fd}.  In order to 
eliminate biases in the underlying event estimation,  an $\eta$-strip of total width $\Delta \eta = 1.6$ 
centered on the jet position was used, with the two highest energy jets in each event excluded~\cite{Cacciari:2010te}.  
In addition, the anti-$k_{\rm T}$ jets were reconstructed based on 
particle flow objects~\cite{particle_flow,CMS-PAS-PFT-10-002} instead of calorimeter-only information.
A good agreement was found with the calorimeter-based, iterative cone algorithm results.

\subsubsection{Simulated data samples}
\label{sec:pythia_samples}

For the analysis of dijet properties in PbPb events, it is crucial to understand how the jet reconstruction
is modified in the presence of the high multiplicity of particles produced in the PbPb underlying event. 
The jet-finding performance was studied using dijets in pp collisions simulated with the
{\sc{pythia}} event generator (version 6.423, tune D6T)~\cite{bib_pythia},
modified for the isospin content of the colliding nuclei \cite{Lokhtin:2005px}.
 In order to enhance the number of Pythia dijets in the momentum range studied, a minimum $\hat{p}_{\rm T}$
selection of 80~\GeVc\ was used. Lower  $\hat{p}_{\rm T}$ selections, as discussed in \cite{Cacciari:2011tm}, were also investigated and found 
to agree with the $\hat{p}_{\rm T} = 80$~\GeVc\ results within uncertainties. The {\sc{pythia}} dijet events were processed with the full detector simulation and analysis chain.
Additional samples
were produced in which the {\sc{pythia}} dijet events were embedded into a minimum bias selection of PbPb events at the raw data level.
For this embedding procedure, both real \PbPb\ data events ({\sc{pythia+data}}), and \PbPb\ events simulated with the 
{\sc{hydjet}} event generator~\cite{Lokhtin:2005px} ({\sc{pythia+hydjet}}) were used.
The {\sc{hydjet}} parameters were tuned to reproduce the total particle multiplicities 
at all centralities and to approximate
the underlying event fluctuations seen in data. The {\sc{hydjet}} events included the simulation
of hard-scattering processes for which radiative parton energy loss was simulated, but 
collisional energy loss was turned off~\cite{Lokhtin:2005px}. Both embedded samples were
propagated through the standard reconstruction and analysis chain.

The {\sc{pythia+data}} sample was used in several ways for studies of calorimeter jets.
First, by matching the same {\sc{pythia}} dijet event
reconstructed with and without the \PbPb\ underlying event, the degradation of the
jet \pt\ and position resolution, the jet \pt\ scale, and the jet-finding efficiency were
determined as a function of collision centrality and jet \pt\ (Section~\ref{sec:jet_performance}).
In addition, {\sc{pythia+data}} events were compared to non-embedded {\sc{pythia}}
for dijet observables such as azimuthal correlations and momentum balance distributions.
Finally, to separate effects due to the medium itself from effects simply due to reconstructing jets
in the complicated environment of the underlying PbPb event, a direct comparison of results for 
{\sc{pythia+data}} and actual data events was made
(Section \ref{sec:dijet_properties}).

The {\sc{pythia+hydjet}} sample was used for studies of track momentum balance and track-jet 
correlations (Sections \ref{sec:track_jet_correlations} and \ref{sec:missingpt}), where access to the full MC particle level (truth) 
information for charged tracks is important for systematic studies. 

\subsubsection{Jet-finding performance}
\label{sec:jet_performance}

\begin{figure}[htb]
\begin{center}
\resizebox{1.0\textwidth}{!}{\includegraphics{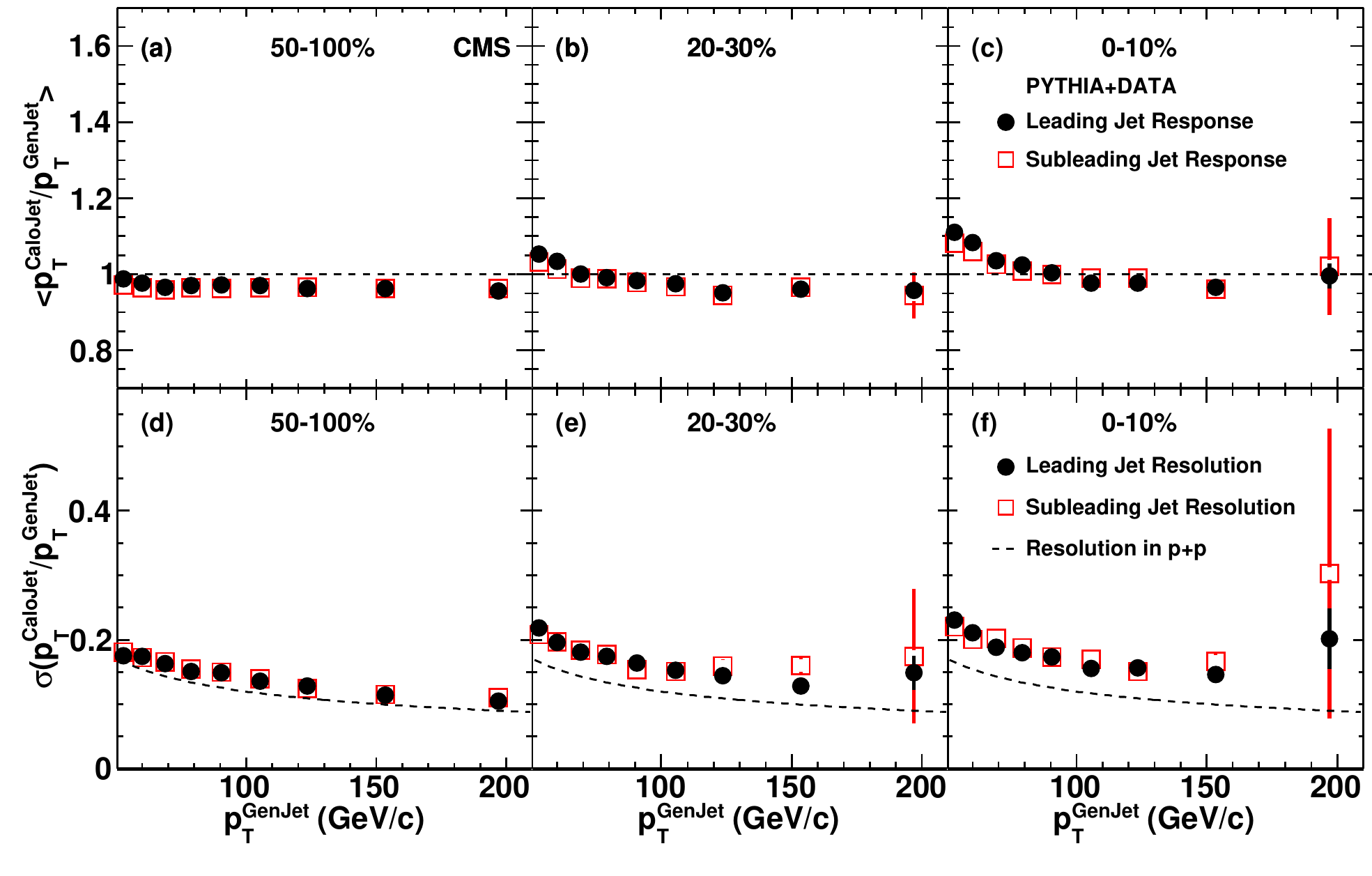}}
\caption{\label{fig:DataMix_JetResponse}
The top row shows the mean of the ratio of reconstructed to generated jet momenta,
$\langle \pt^{\rm CaloJet}/\pt^{\rm GenJet}\rangle$, as a function of $\pt^{\rm GenJet}$, while the bottom
row shows the relative resolution, i.e., the standard deviation of
$\pt^{\rm CaloJet}/\pt^{\rm GenJet}$. The standard \pp\ jet energy corrections are included in $\pt^{\rm CaloJet}$. Filled circles are for the leading jets and open squares are for the subleading jets.  The left, center, and right columns are for jets in
{\sc{pythia+data}} events with centrality 50--100\%, 20--30\%, and 0--10\%, respectively.  On the jet
resolution plots (bottom row), the dashed line is a fit to the leading jet resolution in
\pp\ events. The vertical bars denote the statistical uncertainty.}
\end{center}
\end{figure}

A detailed characterization of the CMS calorimeter jet-finding performance in \pp\ collisions can be found in \cite{bib_jec}.
The dependence of the jet energy scale and of the jet energy resolution on centrality was determined 
using the {\sc{pythia+data}} sample (Fig.~\ref{fig:DataMix_JetResponse}, standard \pp\ jet energy corrections are applied).
In this study, reconstructed jets were matched to the closest generator-level jet in
$\eta$-$\phi$ within a cone of $\Delta R = 0.3$. The residual jet energy scale dependence and the 
relative jet energy resolution are derived from the mean and standard deviation of the approximately 
Gaussian distributions of the ratio of the reconstructed calorimeter jet transverse momentum $\pt^{\rm CaloJet}$ 
and the transverse momentum of jets reconstructed based on event generator level final state particles $\pt^{\rm GenJet}$. 
For peripheral events in the 50--100\% centrality selection, the jet energies are under-corrected by 5\% after 
applying the standard \pp\ jet energy corrections, and the jet energy resolution 
is found to be about 15\% worse than in pp collisions. For the most central events, the large transverse energy per unit area of the underlying event leads to an over-correction 
of low-\pt\ jet energies by up to 10\% and a degradation of the relative resolution by about 30\% to 
$\sigma(\pt^{\rm CaloJet}/\pt^{\rm GenJet}) = 0.16$ at $\pt = 100$~\GeVc. The effect of the underlying event
on the jet angular resolution was also studied. Integrated over jet $\pt > 50$~\GeVc, the angular resolution in
$\phi$ worsens from 0.03 for peripheral events (50--100\%) to 0.04 for central events (0--10\%), while the resolution
in $\eta$ changes from 0.02 to 0.03 over the same centrality range.

\begin{figure}[htb]
\begin{center}
\resizebox{1.0\textwidth}{!}{\includegraphics{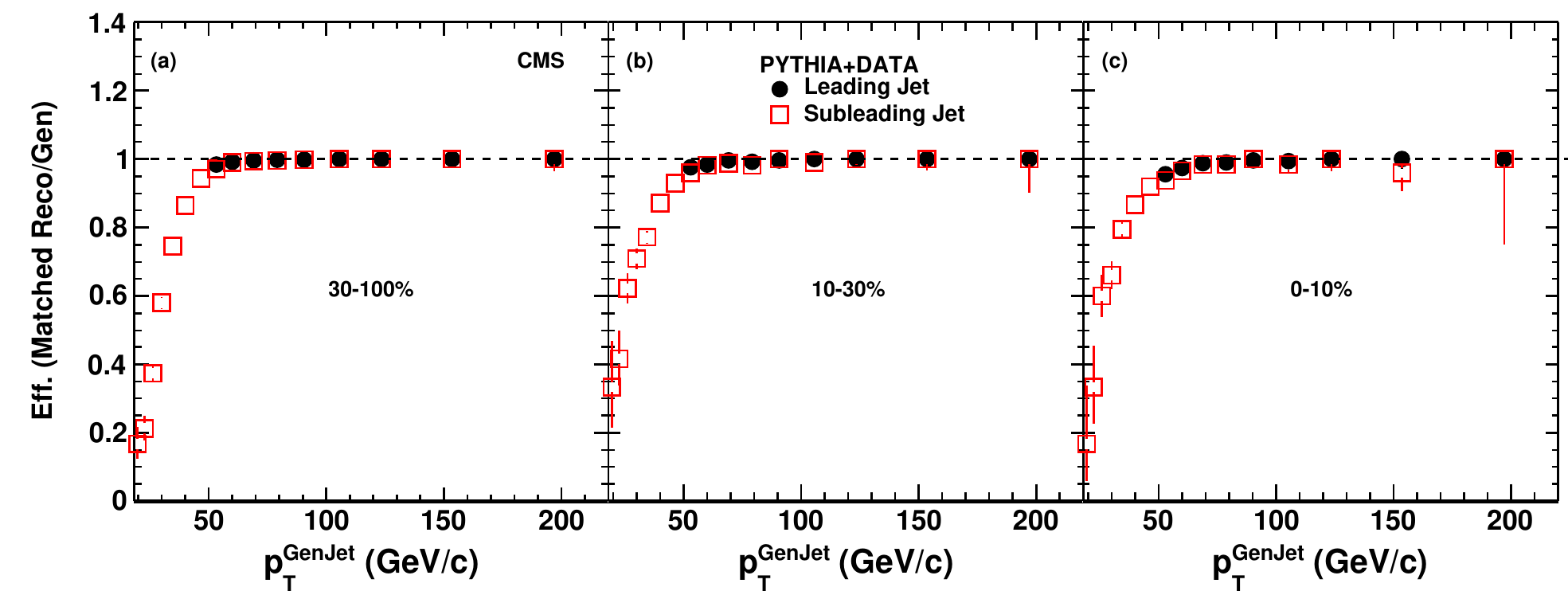}}
\caption{\label{fig:DataMix_JetRecoEff}
Jet reconstruction efficiency as a function of generator level jet $\pt$ for the leading jet (filled circles) and subleading jet (open squares).
From left to right three centrality bins are shown: 30--100\%, 10--30\%, 0--10\%. The vertical bars denote the statistical uncertainty.}
\end{center}
\end{figure}

The jet reconstruction efficiency as a function of jet \pt\ and centrality was extracted from the 
{\sc{pythia+data}} sample as well, 
with the results shown in Fig.~\ref{fig:DataMix_JetRecoEff}. For peripheral events, a jet-finding efficiency of 
95\% was found for a jet 
$\pt = 50$~\GeVc, while for central collisions the efficiency drops to 88\% at the same \pt. Jets with $\pt > 70$~\GeVc are found 
with an efficiency greater than 97\% for all collision centralities. No correction for the inefficiency near the threshold
was applied in the subsequent analysis, as the effects of the reconstruction inefficiency are 
included in the {\sc{pythia+data}} reference analysis.

Finally, the rate of calorimeter jets reconstructed from fluctuations in the underlying event without the presence of a fragmenting $\pt$ parton, so called fake jets,
for the jet selection used in this paper was determined using 
fully simulated 0--10\% central {\sc{hydjet}} events. 
Reconstructed jets in this sample are classified as fake jets if no matching generator-level jet of $\pt > 20$~\GeVc is found within an $\eta$-$\phi$ distance to the reconstructed jet axis smaller than $0.3$.
For leading jets with $p_{\mathrm{T},1}  > 120$~\GeVc, a fake jet fraction of less than 0.02\% is found. 
In events with a $p_{\mathrm{T},1}  > 120$~\GeVc\ leading jet, the fake jet fraction on the 
away-side of the leading jet ($\Delta \phi_{12}> 2\pi/3$) is determined to be 3.5\% for 
reconstructed jets with $p_{\mathrm{T},2} > 50$~\GeVc\ and less than 0.02\% for $p_{\mathrm{T},2}  > 120$~\GeVc.
The effects of the degradation of jet performance in terms of energy scale, resolution, 
efficiency, and fake rate on the dijet observables are discussed in Section~\ref{sec:dijet_properties}.

\section{Results}
\label{sec:results}

The goal of this analysis is to characterize possible modifications of dijet properties as a function
of centrality in \PbPb\ collisions. 
In addition to the standard event selection
of inelastic hadronic collisions and the requirement of a leading 
jet with $p_{\mathrm{T},1} > 120$~\GeVc (Section~\ref{sec:event_selection}),
most of the subsequent analysis required the subleading jet in the event
to have $p_{\mathrm{T},2} > 50$~\GeVc, 
and the azimuthal angle between the leading and subleading jet ($\Delta \phi_{12}$) to be larger than $ 2\pi/3$.
Only jets within $|\eta| < 2$ were considered for the analysis of calorimeter jets in Section~\ref{sec:dijet_properties}.
For a data set of $L_{\rm int} = 6.7~\mu$b$^{-1}$, this selection yields 3514 jet pairs. For studies of correlations
of calorimeter jets with charged particles (Sections~\ref{sec:track_jet_correlations} and \ref{sec:missingpt}), a more restrictive
pseudorapidity selection was applied. 
The analysis was performed mostly in five bins of collision centrality: 0--10\%, 10--20\%, 20--30\%, 30--50\%, and 50--100\%. 

Thus far, no \pp\ reference data exist at the \PbPb\ collision energy of $\rootsNN = 2.76$~TeV.  Throughout the paper, the results obtained 
from \PbPb\ data will be compared to references based on the {\sc{pythia}} and {\sc{pythia+data}} samples described in Section~\ref{sec:pythia_samples}.

For most results, the {\sc{pythia+data}} events will be used for direct comparisons.
To calibrate the performance of {\sc{pythia}} for the observables used in this analysis, the dijet analysis was also performed 
using the anti-$k_{\rm T}$ algorithm on 35~pb$^{-1}$ of \pp\ data at 
$\sqrt{s}=7$~TeV, collected by CMS prior to the heavy ion data taking and compared to {\sc{pythia}} simulations for the same
collision system and energy. The same jet selection criteria used for the 2.76~TeV \PbPb\ data were applied to both \pp\ data and {\sc{pythia}}. 

\subsection{Dijet properties in \pp\ and \PbPb\ data}
\label{sec:dijet_properties}
\begin{figure}[h!]
\begin{center}
\includegraphics[width=1.0\textwidth]{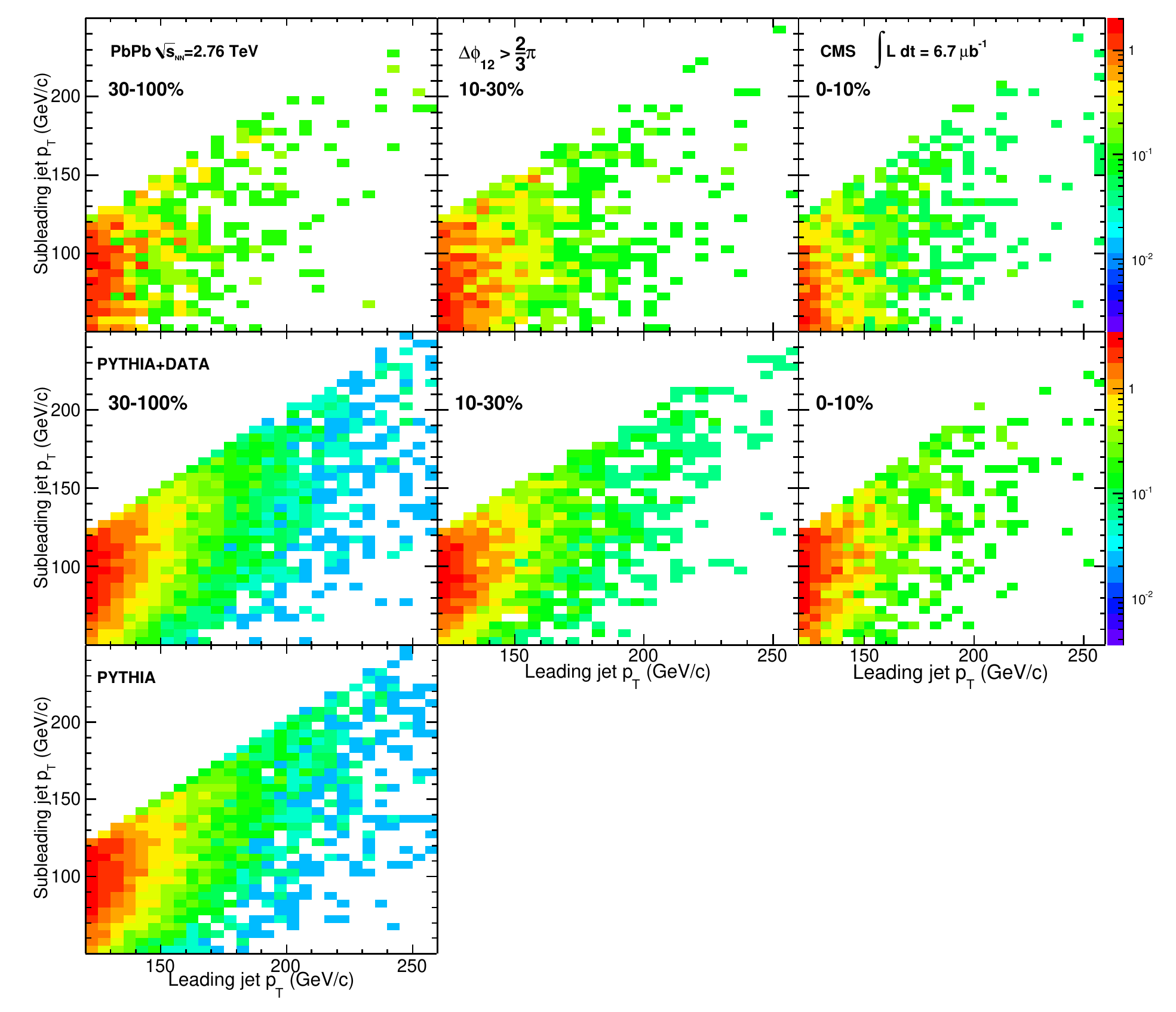}
\caption{
Subleading jet $\pt$ vs. leading jet $\pt$ distributions. The top two rows
show results for centrality
30--100\% (left column),  10--30\% (middle column) and 0--10\% (right column),
for \PbPb\ data (top row) and reconstructed {\sc{pythia}} jets embedded into \PbPb\ data events (middle row).
The panel in the bottom row shows the distribution for
reconstructed jets from {\sc{pythia}} alone.}
\label{fig:et1et2scatter}
\end{center}
\end{figure}

The correlation between the transverse momentum of the reconstructed leading and subleading jets in the
calorimeters is plotted in Fig.~\ref{fig:et1et2scatter}. 
The top row contains \PbPb\ data for peripheral,
mid-central, and central events, the second row shows \pp\ jets simulated by
{\sc{pythia}} and embedded into \PbPb\ data, and the bottom panel shows \pp\ jets from {\sc{pythia}} without
embedding. One can already observe a 
downward shift in the subleading jet \pt\ for the more central \PbPb\ events. In the following discussion, a more 
quantitative and detailed assessment of this phenomenon will be presented.

\subsubsection{Leading jet spectra}
\begin{figure}[h!]
\begin{center}
\includegraphics[width=1.0\textwidth]{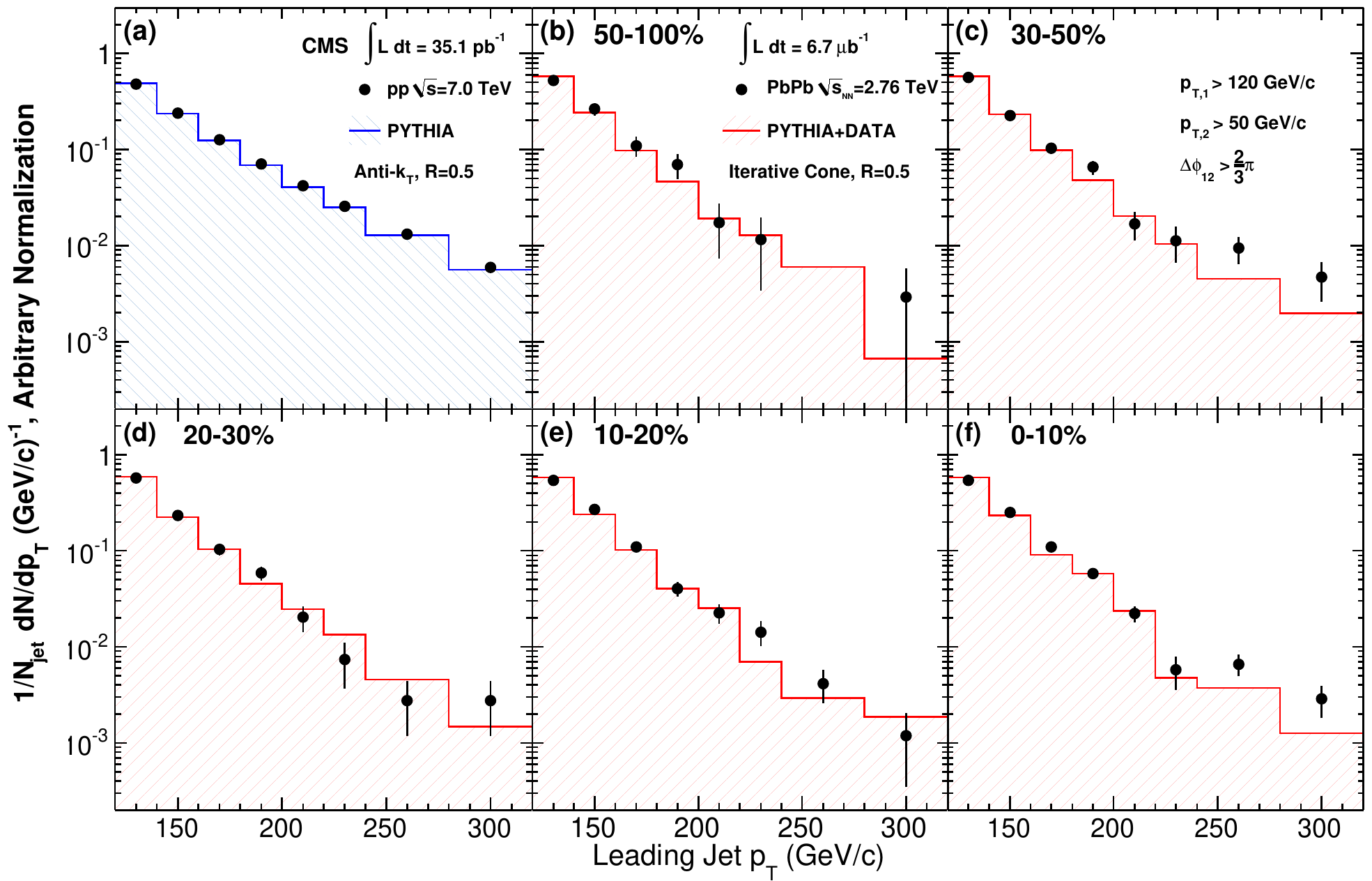}
\caption{Leading jet \pt\ distribution for dijet events with subleading jets of
$p_{\mathrm{T},2} >  50$~\GeVc\ and $\Delta \phi_{12} > 2\pi/3$  for 
 7 TeV pp collisions (a) and 2.76 TeV \PbPb\ collisions in several centrality bins: (b) 50--100\%, (c) 30--50\%, (d) 20--30\%, (e) 10--20\% and (f) 0--10\%.
Data are shown as black points, while the histograms show (a) {\sc{pythia}} events and (b)-(f) {\sc{pythia}} events embedded into \PbPb\ data.  The error bars show the statistical uncertanties.}
\label{fig:LeadingEt}                   
\end{center}
\end{figure}

Figure~\ref{fig:LeadingEt} (a) shows the leading jet \pt\ distributions for 7 TeV pp data and corresponding {\sc{pythia}} simulations. The distribution of leading jet \pt\ for PbPb is shown in Figs.~\ref{fig:LeadingEt} (b)-(f) for five different centrality bins.
The spectra obtained for \PbPb\ data are shown as solid markers, whereas the hatched histograms show the leading 
jet spectrum reconstructed from {\sc{pythia+data}} dijet events.
All spectra have been normalized to unity. The detector-level leading jet spectra in \PbPb\ data and 
the corresponding results for {\sc{pythia+data}} samples show good quantitative agreement
in all centrality bins over the \pt\ range studied.

It is important to note that the jet momentum spectra at detector level presented here have not been corrected for 
smearing due to detector resolution, fluctuations in/out of the jet cone, or underlying event fluctuations.
Therefore, a direct comparison of these spectra to analytical calculations or particle-level generator results is not possible.
For the jet asymmetry and dijet \dphi\ distributions discussed below, the effect of the finite jet energy resolution 
is estimated using the {\sc{pythia+data}} events. 

\subsubsection{Dijet azimuthal correlations}

\begin{figure}[t!]
\begin{center}
\includegraphics[width=1.0\textwidth]{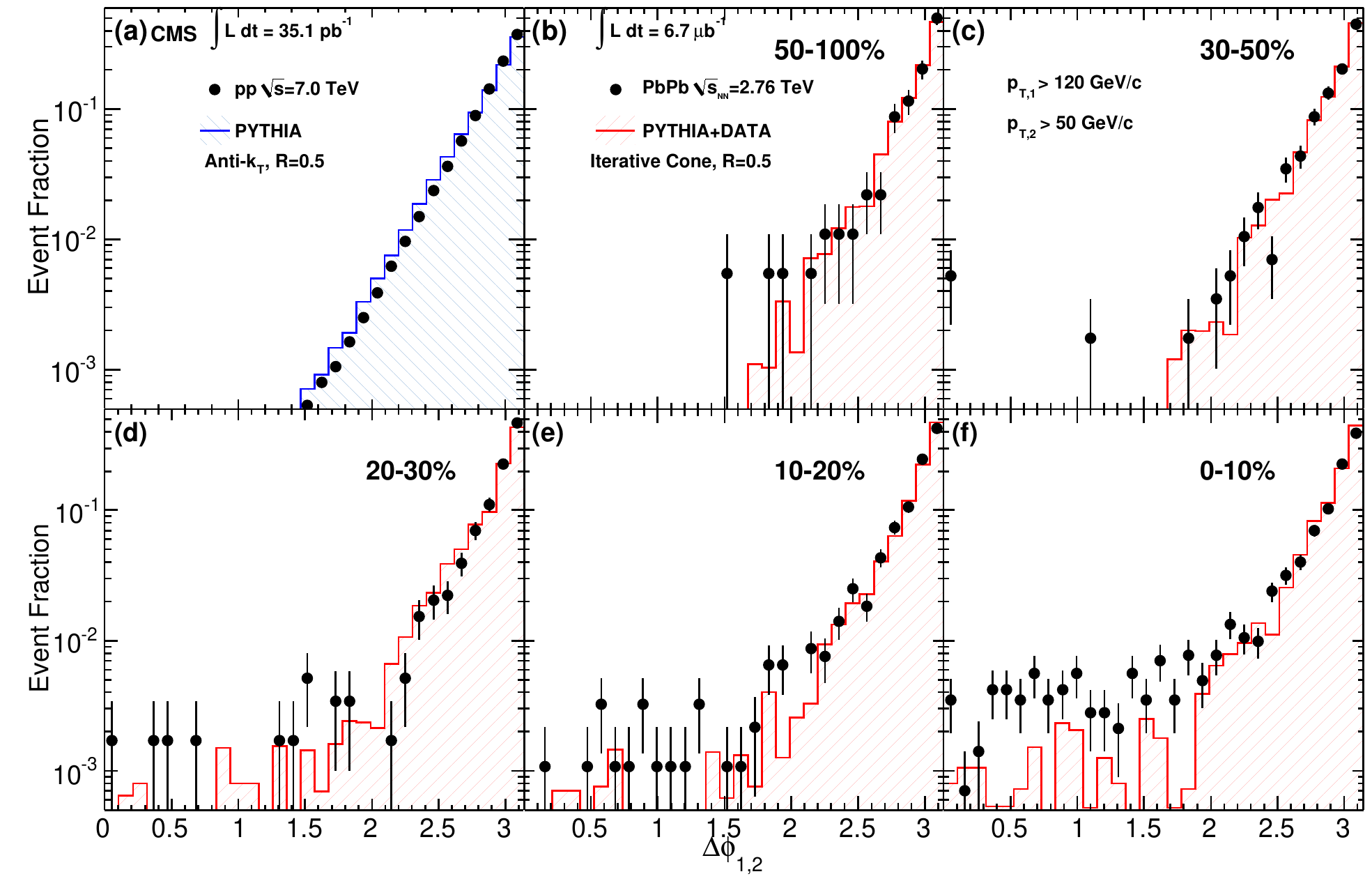}
\end{center}
\caption{$\Delta \phi_{12}$ distributions for leading jets of $p_{\mathrm{T},1} > 120$~\GeVc\ 
with subleading jets of $p_{\mathrm{T},2} >  50$~\GeVc for 
7 TeV pp collisions (a) and 2.76 TeV PbPb collisions in several centrality bins:
(b) 50--100\%, (c) 30--50\%, (d) 20--30\%, (e) 10--20\% and (f) 0--10\%.
Data are shown as black points, while the histograms show (a) {\sc{pythia}} events  and  (b)-(f) {\sc{pythia}} events embedded into PbPb data.  
The error bars show the statistical uncertainties.}
\label{fig:dphi50}
\end{figure}

One possible medium effect on the dijet properties is a change of the back-to-back alignment of 
the two partons. This can be studied using the event-normalized differential dijet distribution, 
($1/N$)($dN / d\Delta \phi_{12}$), 
versus $\Delta \phi_{12}$. 
Figure~\ref{fig:dphi50} shows distributions of $\Delta \phi_{12}$ between leading and subleading jets which pass the respective \pt selections.
In Fig.~\ref{fig:dphi50} (a), the dijet $\Delta \phi_{12}$ distributions are plotted for 7~TeV \pp\ data in comparison to 
the corresponding {\sc{pythia}} simulations using the anti-$k_{\rm T}$ algorithm for jets based on calorimeter information. 
{\sc{pythia}} provides a good description of the experimental data, with slightly larger tails seen in the {\sc{pythia}} simulations. 
A recent study of azimuthal correlations in \pp\ collisions at 7~TeV can be found in~\cite{Collaboration:2011zj}.
For the {\sc{pythia}} comparison to \PbPb\ results at $\rootsNN = 2.76$~TeV, this discrepancy 
seen in the higher energy pp comparison is included in the systematic uncertainty estimation. 
It is important to note that the {\sc{pythia}} simulations include events with more than two jets, which 
provide the main contribution to events with large momentum imbalance or 
$\Delta \phi_{12}$ far from $\pi$.

Figures~\ref{fig:dphi50}~(b)-(f) show the dijet $\Delta \phi_{12}$ distributions for \PbPb\ data in five centrality bins,
compared to {\sc{pythia+data}} simulations.  The distributions for the
four more peripheral bins are in good 
agreement with the {\sc{pythia+data}} reference, especially for $\dphi_{12} \gtrsim 2$. 
The three centrality bins spanning 0--30\% show an excess of events with azimuthally misaligned dijets 
($\dphi_{12} \lesssim 2$), compared with more
peripheral events.  A similar trend is seen for the {\sc{pythia+data}} simulations, although the 
fraction of events with  azimuthally misaligned dijets is smaller in the simulation. The centrality dependence 
of the azimuthal correlation in {\sc{pythia+data}} can
be understood as the result of the increasing fake-jet rate and the
drop in jet reconstruction efficiency near the 50~\GeVc\ threshold from 95\% for
peripheral events to 88\% for the most central events. In \PbPb\ data, 
this effect is magnified since low-\pt away-side jets can undergo a sufficiently large energy loss to fall below the 50~\GeVc\ 
selection criteria.

\begin{figure}[h!]
\begin{center}
\includegraphics[width=0.48\textwidth]{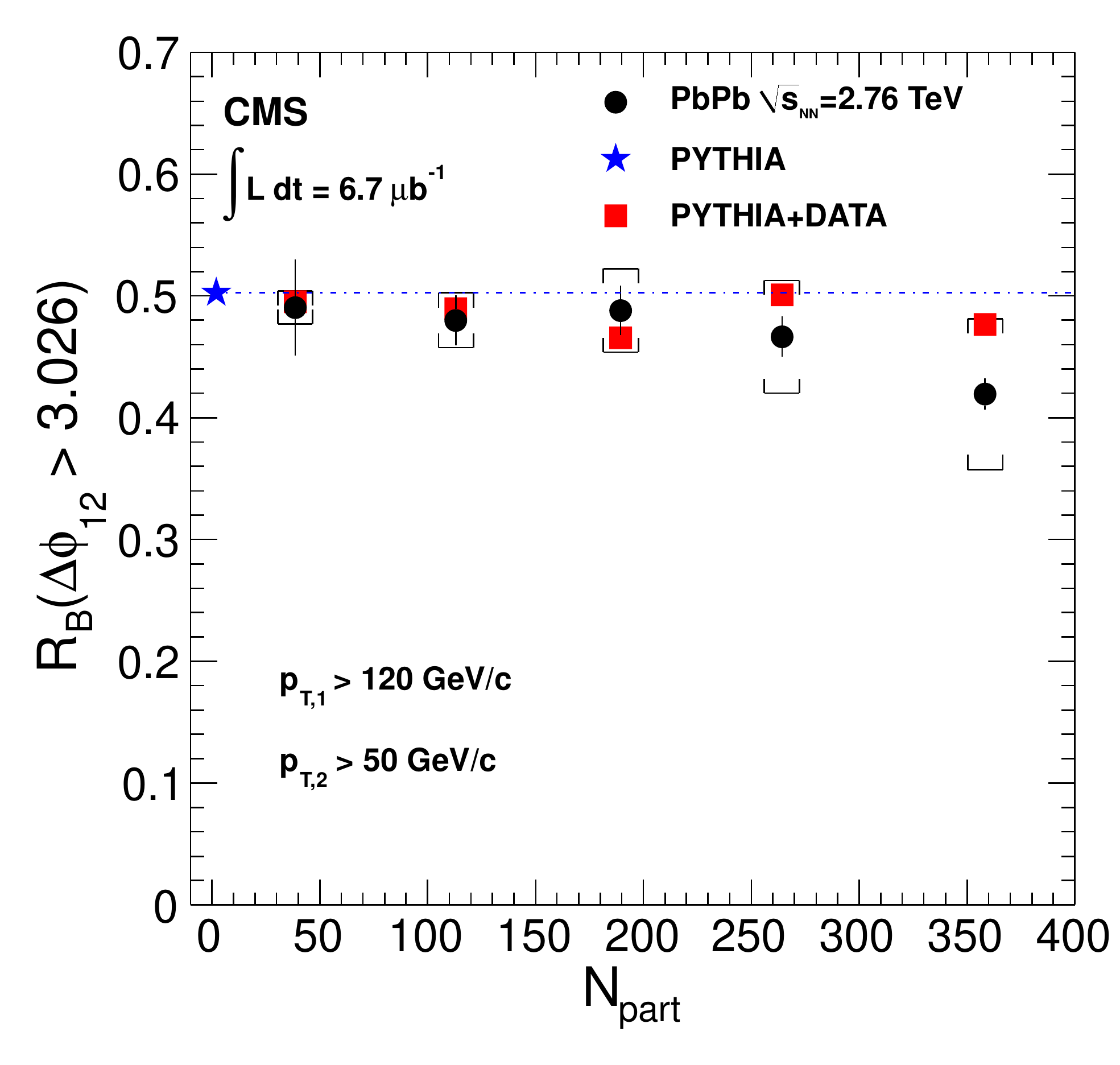}
\caption{
Fraction of events with $\Delta\phi_{12}>3.026$  
as a function of \npart, among events with $p_{\mathrm{T},1} > 120$~\GeVc\  
and $p_{\mathrm{T},2} > 50$~\GeVc.
The result for reconstructed {\sc{pythia}} dijet events (blue filled star)
 is plotted at \npart\ = 2.  The other points (from left to right) correspond to centrality bins of 
50--100\%, 30--50\%, 20--30\%, 10--20\%, and 0--10\%. 
The red squares are for reconstruction of {\sc{pythia+data}} events and the filled
circles are for the \PbPb\ data, with statistical (vertical bars) and systematic (brackets) 
uncertainties. }
\label{fig:rbdphi}                  
\end{center}
\end{figure}

Furthermore, a reduction of the fraction of back-to-back jets above $\dphi_{12}\gtrsim 3$  is observed
for the most central bin.
This modification of the $ \dphi_{12}$ distribution as a function of centrality can be quantified 
using the fraction $R_B$ of dijets with $\Delta\phi_{12} > 3.026$ , as plotted 
in Fig.~\ref{fig:rbdphi}, for $p_{\mathrm{T},1} > 120$~\GeVc and $p_{\mathrm{T},2} > 50$~\GeVc.
The threshold of 3.026  corresponds to the median of the $\Delta\phi_{12}$ distribution for {\sc{pythia}} (without embedding).  
The results for both the \PbPb\ data and {\sc{pythia+data}} dijets are shown 
as a function of the reaction centrality, given by the number of participating nucleons, 
\npart, as described in Section~\ref{sec:centrality}. 
This observable is not sensitive to the shape of the tail at $\Delta\phi_{12} <2$  seen in Fig.~\ref{fig:dphi50}, but 
can be used to measure small changes in the back-to-back correlation between dijets.
A decrease in the fraction of back-to-back jets in \PbPb\ data is seen compared to the pure {\sc{pythia}} simulations.
Part of the observed change in $R_B(\Delta \phi)$ with centrality is explained by the decrease in jet azimuthal angle 
resolution from $\sigma_\phi = 0.03$ in peripheral events to $\sigma_\phi = 0.04$ in central events,
due to the impact of fluctuations in the \PbPb\ underlying event. This effect is demonstrated by the 
comparison of {\sc{pythia}} and {\sc{pythia+data}} results. 
The difference between the \pp\ and {\sc{pythia+data}} resolutions was used for the uncertainty 
estimate, giving the dominant contribution to the systematic uncertainties, shown as brackets in  Fig.~\ref{fig:rbdphi}.

\subsubsection{Dijet momentum balance}
\label{sec:asymmetry}
To characterize the dijet momentum balance (or imbalance) quantitatively, we use the asymmetry ratio,
\begin{equation}
\label{eq:aj} 
A_J = \frac{p_{\mathrm{T},1}-p_{\mathrm{T},2}}{p_{\mathrm{T},1}+p_{\mathrm{T},2}}~,
\end{equation} 
where the subscript $1$ always refers to the leading jet, so that $A_J$ is 
positive by construction. 
The use of $A_J$ removes uncertainties due to possible constant shifts of the jet energy scale. 
It is important to note that the subleading jet $p_{\mathrm{T},2} > 50$~\GeVc\ selection imposes
a $p_{\mathrm{T},1}$-dependent limit on the magnitude of \AJ. For example, for the most frequent leading
jets near the 120~\GeVc\ threshold, this 
limit is $\AJ < 0.41$, while the largest possible $A_J$ for the present dataset is 0.7 for 
300~\GeVc\ leading jets. Dijets in which the subleading jet is lost below the 50~\GeVc\ threshold 
are not included in the $\AJ$ calculation.

\begin{figure}[t!]
\begin{center}
\includegraphics[width=1.0\textwidth]{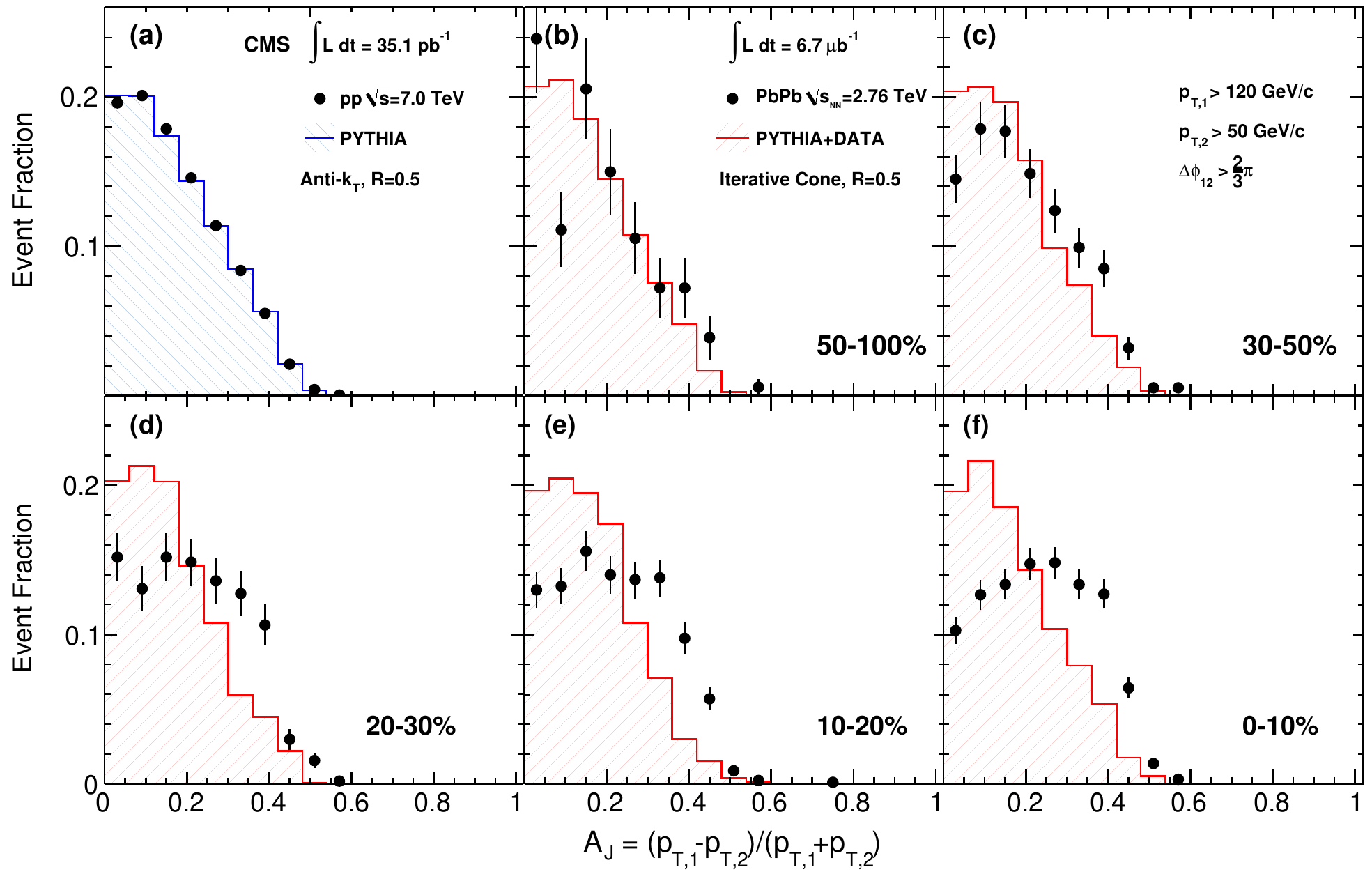}
\caption{Dijet asymmetry ratio, $A_{J}$, for leading jets of $p_{\mathrm{T},1}> $ 120~\GeVc, subleading jets of $p_{\mathrm{T},2}> $50~\GeVc\, and $\Delta\phi_{12}>2\pi/3$  for 7 TeV pp collisions (a) and 2.76 TeV \PbPb\ collisions in several centrality bins:  (b) 50--100\%, (c) 30--50\%, (d) 20--30\%, (e) 10--20\% and (f) 0--10\%.
Data are shown as black points, while the histograms show  (a) {\sc{pythia}} events and  (b)-(f) {\sc{pythia}} events embedded into \PbPb\ data.  
The error bars show the statistical uncertainities.}
\label{fig:JetAsymm}                  
\end{center}
\end{figure}

In Fig.~\ref{fig:JetAsymm} (a),  the $A_J$ dijet asymmetry observable calculated 
by {\sc{pythia}} is compared to \pp\  data at $\sqrt{s}$ = 7~TeV. Again, data and event generator
are found to be in excellent agreement. This observation, as well as the good agreement
between {\sc{pythia+data}} and the most peripheral \PbPb\ data shown in 
Fig.~\ref{fig:JetAsymm} (b), suggests that {\sc{pythia}}
at $\sqrt{s}$ = 2.76~TeV can serve as a good reference for the dijet imbalance analysis
in \PbPb\ collisions.

The centrality dependence of $A_J$ for \PbPb\ collisions can be seen in
Figs.~\ref{fig:JetAsymm}~(b)-(f), in comparison to {\sc{pythia+data}} simulations.
Whereas the dijet angular correlations show only a small dependence on collision
centrality, the dijet momentum balance exhibits a dramatic change in shape
for the most central collisions. In contrast, the {\sc{pythia}} simulations only
exhibit a modest broadening, even when embedded in the highest multiplicity 
\PbPb\ events. 

Central \PbPb\ events show a significant deficit of events in which the momenta of leading and subleading jets are balanced and a
significant excess of unbalanced pairs. The large excess of
unbalanced compared to balanced dijets explains why this effect was
apparent even when simply scanning event displays (see
Fig.~\ref{fig:eventDisplay}). The striking momentum imbalance
is also confirmed  when studying high-\pt\ tracks associated with 
leading and subleading jets, as will be shown in Section~\ref{sec:track_jet_correlations}.
It is consistent with a degradation of the parton energy, or jet quenching, in the 
medium produced in central \PbPb\ collisions.

\begin{figure}[h!]
\begin{center}
\includegraphics[width=0.48\textwidth]{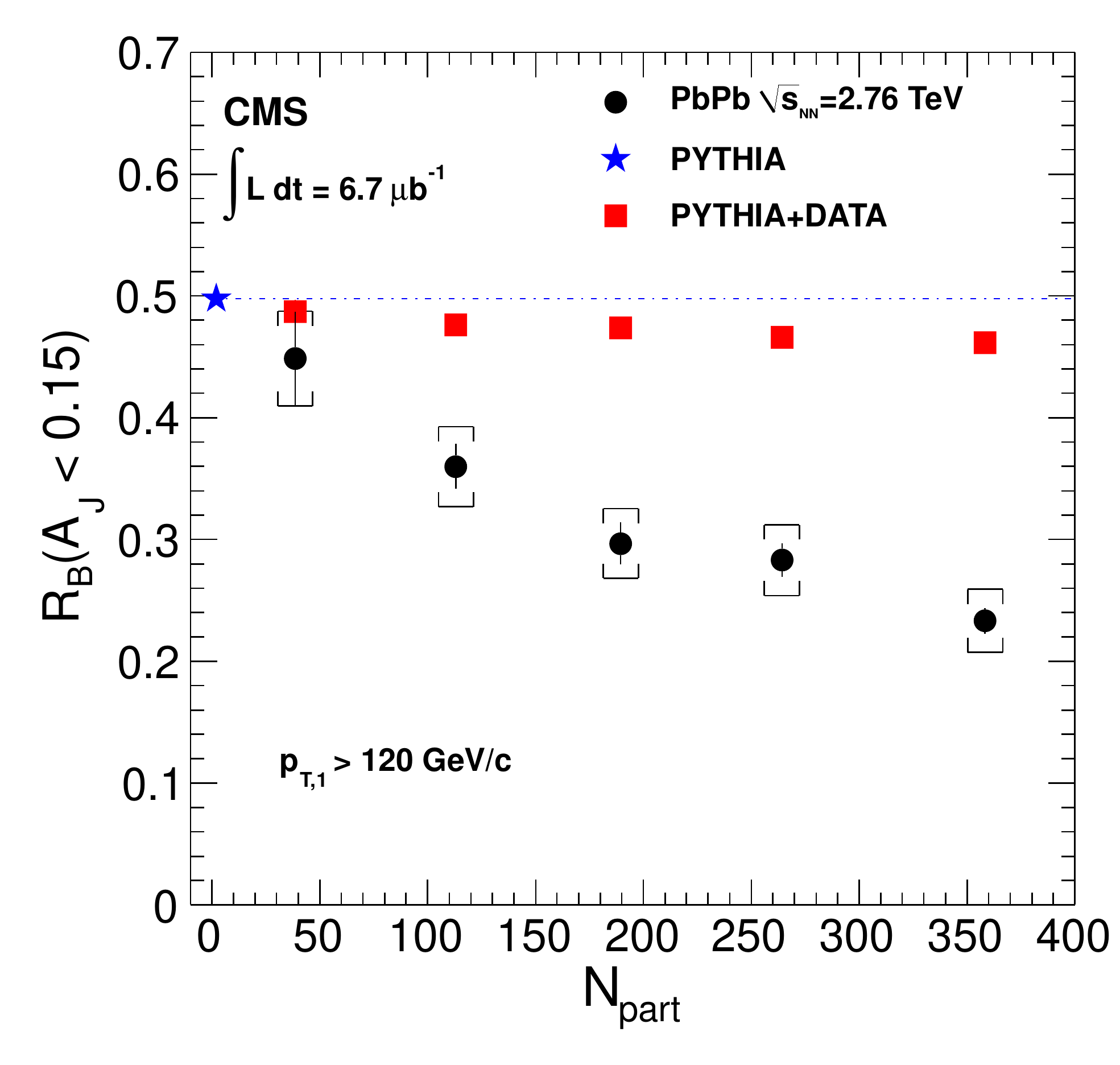}
\caption{
Fraction of all events with a
leading jet with $p_{\mathrm{T},1} > 120$~\GeVc\ for which a subleading jet
with $A_J<0.15$ and $\dphi_{12} > 2\pi/3$
was found, as a function of \npart.
The result for reconstructed {\sc{pythia}} dijet events (blue filled star)
is plotted at \npart\ = 2.  The other points (from left to right) correspond to centrality bins of 
50--100\%, 30--50\%, 20--30\%, 10--20\%, and 0--10\%. 
The red squares are for reconstruction of {\sc{pythia+data}} events and the filled
circles are for the \PbPb\ data, with statistical
(vertical bars) and systematic (brackets) uncertainties. }
\label{fig:rbAJ}
\end{center}
\end{figure}

The evolution of the dijet momentum balance illustrated in Fig.~\ref{fig:JetAsymm} can be explored 
more quantitatively by studying the fraction of balanced jets in the \PbPb\ events.
The balanced fraction, $R_B(\AJ < 0.15)$, is plotted as a function of collision centrality (again in terms of \npart) in 
Fig.~\ref{fig:rbAJ}. It is defined as the fraction of all events with a leading jet having $p_{\mathrm{T},1} > 120$~\GeVc\
for which a subleading partner with $\AJ < 0.15$ and $\dphi_{12} > 2\pi/3$
is found.  Since $R_B(\AJ < 0.15)$ is calculated as the fraction of all events with $p_{\mathrm{T},1} > 120$~\GeVc, 
it takes into account the rate of apparent ``mono-jet'' events, where the subleading partner is removed by the 
\pt\ or \dphi\ selection. 

The \AJ\ threshold of 0.15 corresponds to the median of the \AJ\ distribution
for pure {\sc{pythia}} dijet events passing the criteria used for Fig.~\ref{fig:JetAsymm}.
By definition, the fraction $R_B(\AJ < 0.15)$ of balanced jets in {\sc{pythia}}
is therefore 50\%, which is plotted as a dashed line
in Fig.~\ref{fig:rbAJ}. As will be discussed in Section~\ref{sec:missingpt}, a third jet having a significant impact on the dijet imbalance is present in most of the large-\AJ\ events in {\sc{pythia}}.

The change in jet-finding performance from high to low \pt, discussed in Section~\ref{sec:jet_performance}, leads to only a small decrease in the fraction of balanced jets, of less than 5\% for central  {\sc{pythia+data}}  dijets. In contrast, the \PbPb\ data show a rapid 
decrease in the fraction of balanced jets with collision centrality.
While the most peripheral selection shows a fraction of balanced jets of close
to 45\%, this fraction drops  by close to a factor of two for the most central
collisions.  This again suggests that the passage of hard-scattered partons through the
environment created in \PbPb\ collisions has a significant impact on
their fragmentation into final-state jets.

The observed change in the fraction of balanced jets as a function of centrality, 
shown in Fig.~\ref{fig:rbAJ}, is far bigger than the estimated systematic uncertainties, 
shown as brackets. The main contributions to the systematic uncertainties 
include the uncertainties on jet energy scale and resolution, jet reconstruction 
efficiency, and the effects of underlying event subtraction. The uncertainty in the 
subtraction procedure is estimated based on the difference between pure {\sc{pythia}}
and {\sc{pythia+data}} simulations.  For central events, the 
subtraction procedure contributes the biggest uncertainty to $R_B(\AJ)$, of
close to 8\%. The uncertainty on the residual jet energy scale was estimated based 
on the results shown in the top row of
Fig.~\ref{fig:DataMix_JetResponse}. The full difference between the 
observed residual correction and unity, added in quadrature with 
the systematic uncertainty obtained for \pp\ \cite{CMS-PAS-JME-10-010}, 
was used as the systematic uncertainty on
the jet \pt\ and propagated to $R_B(\AJ)$. For the 
jet \pt\ resolution uncertainty, the full difference of the {\sc{pythia+data}} result
to the \pp\ resolution, as shown in Fig.~\ref{fig:DataMix_JetResponse}~(bottom),
was used as an uncertainty estimate for the \PbPb\ jet \pt\ resolution.
The uncertainties in jet energy scale and jet resolution contribute 5\% and 6\%, respectively, 
to the 11\% total systematic uncertainty in central events. 
For peripheral events, the total uncertainty drops to 9\%, mostly due to the smaller
uncertainty related to the \PbPb\ background fluctuations for lower multiplicity 
events.

\subsubsection{Leading jet \pt\ dependence of dijet momentum imbalance}

\begin{figure}[h!]
\begin{center}
\includegraphics[width=1.0\textwidth]{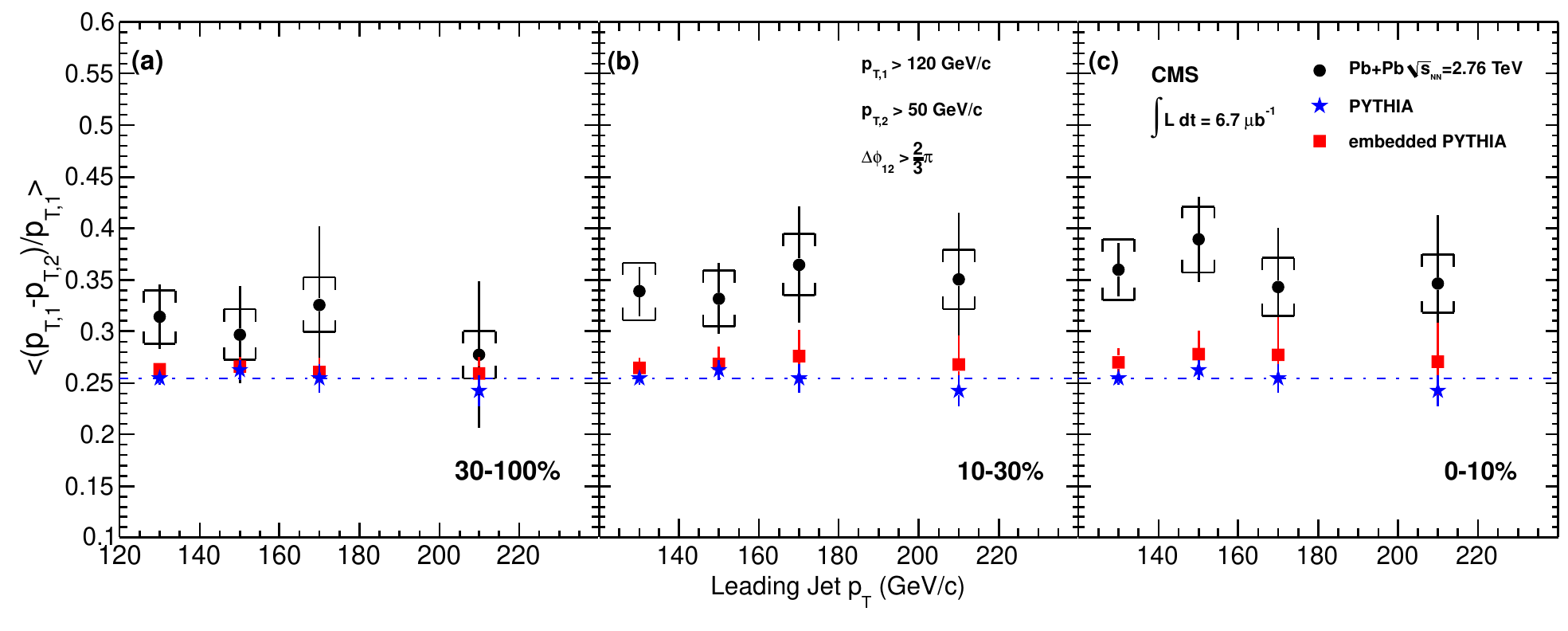}
\caption{
Mean value of the fractional imbalance $(p_{\mathrm{T},1} - p_{\mathrm{T},2})/p_{\mathrm{T},1}$ as a function of leading jet \pt for three centrality bins. 
The \PbPb\ data are shown as circles with vertical bars and brackets indicating the statistical and systematic
uncertainties, respectively.  Results for {\sc{pythia}} are shown with blue stars, and {\sc{pythia+data}} with red squares. The dot-dashed line to guide the eye is drawn at the value for pure {\sc pythia} for the lowest \pt bin.}
\label{deltaPt}                  
\end{center}
\end{figure}

The dependence of the jet modification on the leading jet momentum can be studied using the fractional imbalance 
$\Delta p_{\mathrm{T}_{rel}} = (p_{\mathrm{T},1} - p_{\mathrm{T},2})/p_{\mathrm{T},1}$. 
The mean value of this fraction is presented as a function of $p_{\mathrm{T},{1}}$  
in Fig.~\ref{deltaPt} for three bins of collision centrality, 30--100\%, 10--30\% and 0--10\%.
{\sc{pythia}} is shown as stars, {\sc{pythia+data}} simulations are shown as squares, while the 
data are shown as circles. Statistical and systematic uncertainties are plotted as error bars and brackets, respectively.
The dominant contribution to the systematic uncertainty comes from the observed \pt\ dependence
of the residual jet energy correction in \PbPb\ events (6\%  out of a total systematic uncertainty of 8\%). 
The jet energy resolution and underlying event subtraction uncertainties contribute about 4\% each.

The fractional imbalance exhibits several important features: 
the imbalance seen in \PbPb\ data grows with  collision centrality and reaches a much  larger 
value than in {\sc{pythia}} or {\sc{pythia+data}}. 
In addition, the effect is clearly visible even for the highest-\pt\  jets observed in the data set, demonstrating that
the observed dijet imbalance is not restricted to the threshold region in our leading jet selection. 
Within the present uncertainties, the $p_{\mathrm{T},1}$ dependence of the excess imbalance above the {\sc{pythia}} prediction is compatible with either a constant difference or a constant fraction of $p_{\mathrm{T},1}$.

The main contributions to the systematic uncertainty in $(p_{\mathrm{T},1} - p_{\mathrm{T},2})/p_{\mathrm{T},1}$
are the uncertainties in the \pt-dependent residual energy scale (based on results shown in the top 
row of Fig.~\ref{fig:DataMix_JetResponse}), and the centrality-dependent difference
observed between {\sc{pythia}} and {\sc{pythia+data}} seen in Fig.~\ref{deltaPt}. As before,
the uncertainty on the residual jet energy scale was estimated
using the full difference between the observed residual correction and unity, and also assuming that within these limits the low-\pt\ and high-\pt\ response could vary independently.

\subsection{Track-jet correlations}
\label{sec:track_jet_correlations}

The studies of calorimeter jets show a strong change of the jet momentum balance 
as a function of collision centrality. This implies a corresponding modification 
in the distribution of jet fragmentation products, with energy being either 
transported out of the cone area used to define the jets, or to low-momentum 
particles which are not measured in the calorimeter jets. The CMS calorimeter is less sensitive to these low momentum particles, or  they do not reach the calorimeter surface.
Information about changes to the effective fragmentation pattern as a function of \AJ\ 
can be obtained from track-jet correlations. For this analysis, {\sc{pythia+hydjet}} simulations are
used as MC reference, to allow full access to MC truth (i.e., the output of  the generator) information for tracks in the dijet signal and in  the \PbPb\ underlying event. The event selection for {\sc{pythia+hydjet}} was  based on
reconstructed calorimeter jet information, as  for the previous studies.

To derive the associated track spectrum for a given jet selection in data, the \pt\ distribution of tracks inside a 
ring of radius $\Delta R=\sqrt{\Delta\phi^2+\Delta \eta^2}$ and width of 0.08 around the jet axes was summed over all selected jets. 
The 
contribution of tracks from the underlying event, not associated with the jet, was estimated 
by summing the track \pt\ distributions using an equal-size ring that was reflected around $\eta=0$,
but at the same $\phi$ coordinate as the individual jet. For this procedure,
jets in the region $|\eta| < 0.8$ were excluded and only ring-radii up to 
$\Delta R=0.8$ around the jet axes were considered, to avoid overlap between the signal jet region and the region 
used for background estimation. In addition, jets in the region $|\eta| > 1.6$ were
excluded to ensure the 0.8 radius rings would lie within the tracker acceptance.
Statistical fluctuations in the underlying event limit this procedure to tracks with transverse momenta $\pt > 1$~\GeVc.

The summed \pt\ spectra from the jet regions and the underlying event regions were then subtracted, 
yielding the momentum distribution of charged tracks associated with the jets
as a function of $\Delta R$. 

\begin{figure}[t!]
  \begin{center}
    \includegraphics[width=1.0\textwidth]{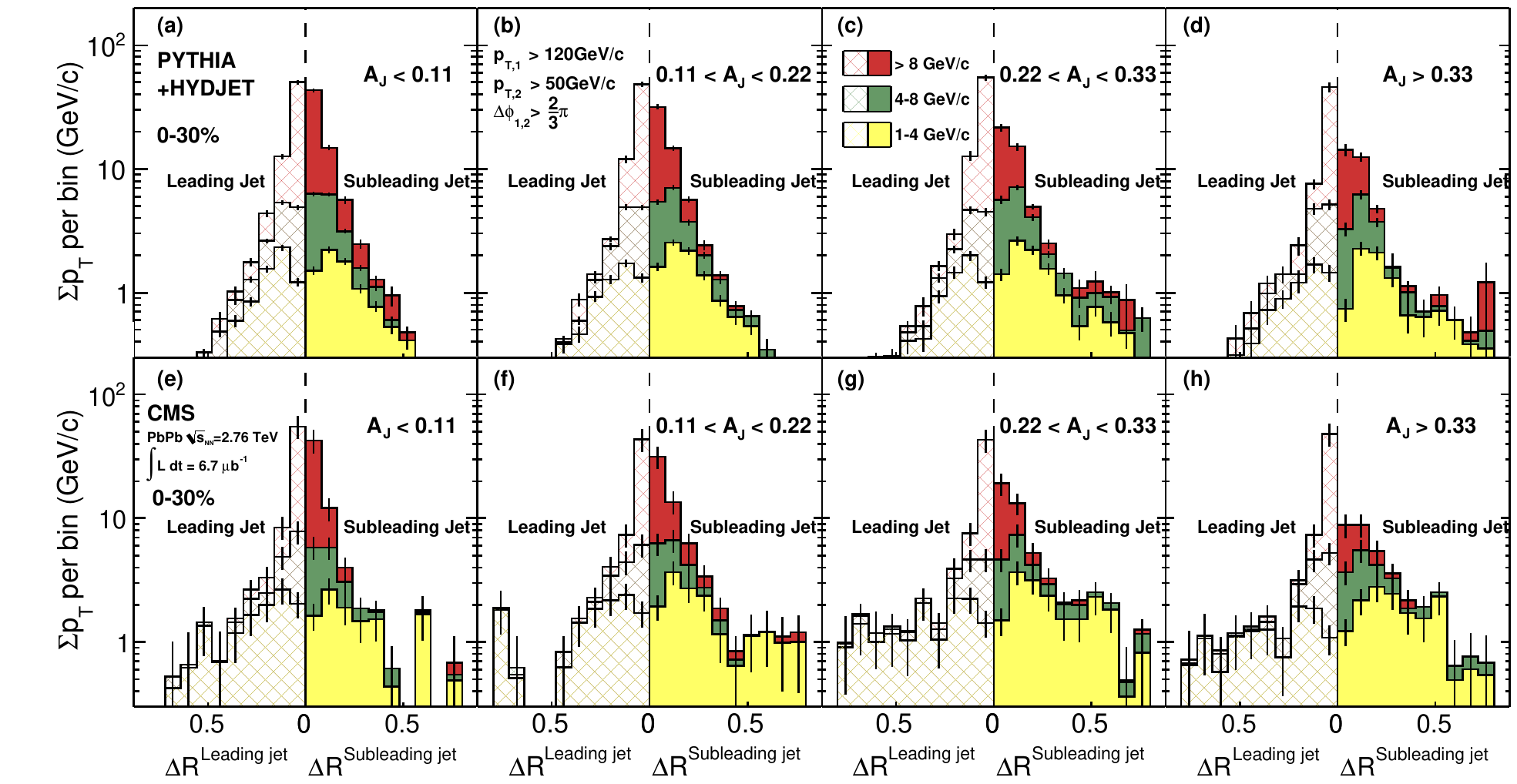}
    \caption{ Distribution of the transverse momentum sum of tracks for three \pt\ ranges, as a function of the distance
$\Delta R$ to the leading and subleading jet axes. Results for the 0--30\% centrality selection are shown for
{\sc{pythia}}+{\sc{hydjet}} (upper row) and \PbPb\ data (lower row).
For each figure, the requirements on the dijet asymmetry $A_J$ are given.
Note that events with $A_J > 0.22$ are much rarer in the {\sc{pythia}}+{\sc{hydjet}} sample than in the data. 
Vertical bars are statistical and systematic uncertainties,
combined in quadrature, the systematic contributions being
20\%, independent of the bin.
}
    \label{fig:dRandPtvsAJcut}
  \end{center}
\end{figure}

The resulting distributions of associated track momentum as a function of track \pt\ and $\Delta R$
are presented in Fig.~\ref{fig:dRandPtvsAJcut} for four selections in dijet asymmetry,
from $A_J < 0.11$ (left) to $A_J > 0.33$ (right). For both data and {\sc{pythia+hydjet}} results,
the jet selections and \AJ\ values are based on the reconstructed calorimeter jet momenta
(Section~\ref{sec:jet_reconstruction}) in order to have consistent event selections for comparison.
The middle bin boundary ($\AJ = 0.22$) corresponds to the median of the \AJ\ distribution for 
the 0--30\% central \PbPb\ events shown here.  
The top row shows the results for {\sc{pythia+hydjet}} simulations.
The track results shown for the {\sc{pythia+hydjet}} simulations
were found using the known (``truth'') values of the track momenta from the embedded {\sc{pythia}} events.
The bottom row presents results for \PbPb\ data.
The track results shown for \PbPb\ data were corrected for tracking efficiency
and fake rates using corrections that were derived from {\sc{pythia+hydjet}} simulations and 
from the reconstruction of single tracks embedded in data.
In each panel, the area of each colored region in 
\pt\ and $\Delta R$  corresponds to the total transverse momentum per event
carried by tracks in this region. 

For the balanced-jet selection, $\AJ < 0.11$, one sees qualitative agreement in the leading 
and subleading jet momentum distributions between {\sc{pythia}}+{\sc{hydjet}} (top) and data (bottom).
In data and simulation, most of the leading and subleading jet momentum is carried by tracks 
with $\pt > 8$~\GeVc, with the data tracks having a slightly narrower $\Delta R$ distribution. 
A slightly larger fraction of the momentum for the subleading jets is carried by 
tracks at low \pt\ and $\Delta R > 0.16$ (i.e., beyond the second bin) in the data.

Moving towards larger dijet imbalance, the major fraction of the leading jet momentum 
continues to be carried by high-\pt\ tracks in data and simulation. 
For the $\AJ > 0.33$ selection, it is important to recall that less than 10\% of all {\sc{pythia}}
dijet events fall in this category, and, as will be discussed in Section~\ref{sec:missingpt},
those that do are overwhelmingly 3-jet events. 

While the overall change found in the leading jet shapes as a function of \AJ\ is small, 
a strong modification of the track momentum composition of the subleading jets is seen, confirming 
the calorimeter determination of the dijet imbalance.  The biggest difference between data and 
simulation is found for tracks with $\pt < 4$~\GeVc.
For {\sc{pythia}}, the momentum in the subleading jet carried by these tracks is small 
and their radial distribution is nearly
unchanged with \AJ. 
However, for data, the relative contribution of low-\pt\ tracks
grows with \AJ, and an increasing fraction of those tracks is observed at
large distances to the jet axis, extending out to $\Delta R=0.8$ (the largest angular distance to the jet in this study).

The major systematic uncertainties 
for the track-jet correlation measurement
come from the $\pt$-dependent uncertainty in the track reconstruction efficiency.  
The algorithmic track reconstruction efficiency, which averages 70\% over the 
$\pt > 0.5$~\GeVc\  and $|\eta| < 2.4$ range included in this study, 
was determined from an independent {\sc{pythia+hydjet}} sample, and from simulated tracks embedded in 
data. Additional uncertainties are introduced
by the underlying event subtraction procedure. The latter was studied by comparing 
the track-jet correlations seen in pure {\sc{pythia}} dijet events for generated particles
with those seen in {\sc{pythia+hydjet}} events after reconstruction and background subtraction.
The size of the background subtraction systematic uncertainty was further cross-checked in
data by repeating the procedure for random ring-like regions in 0--30\% central minimum bias events.
In the end, an overall systematic uncertainty of 20\% per bin was assigned.
This uncertainty is included in the combined statistical and systematic uncertainties shown in Fig.~\ref{fig:dRandPtvsAJcut}.

\subsection{Overall momentum balance of dijet events}
\label{sec:missingpt}

\begin{figure}[th!]
  \begin{center}
    \includegraphics[width=1.0\textwidth]{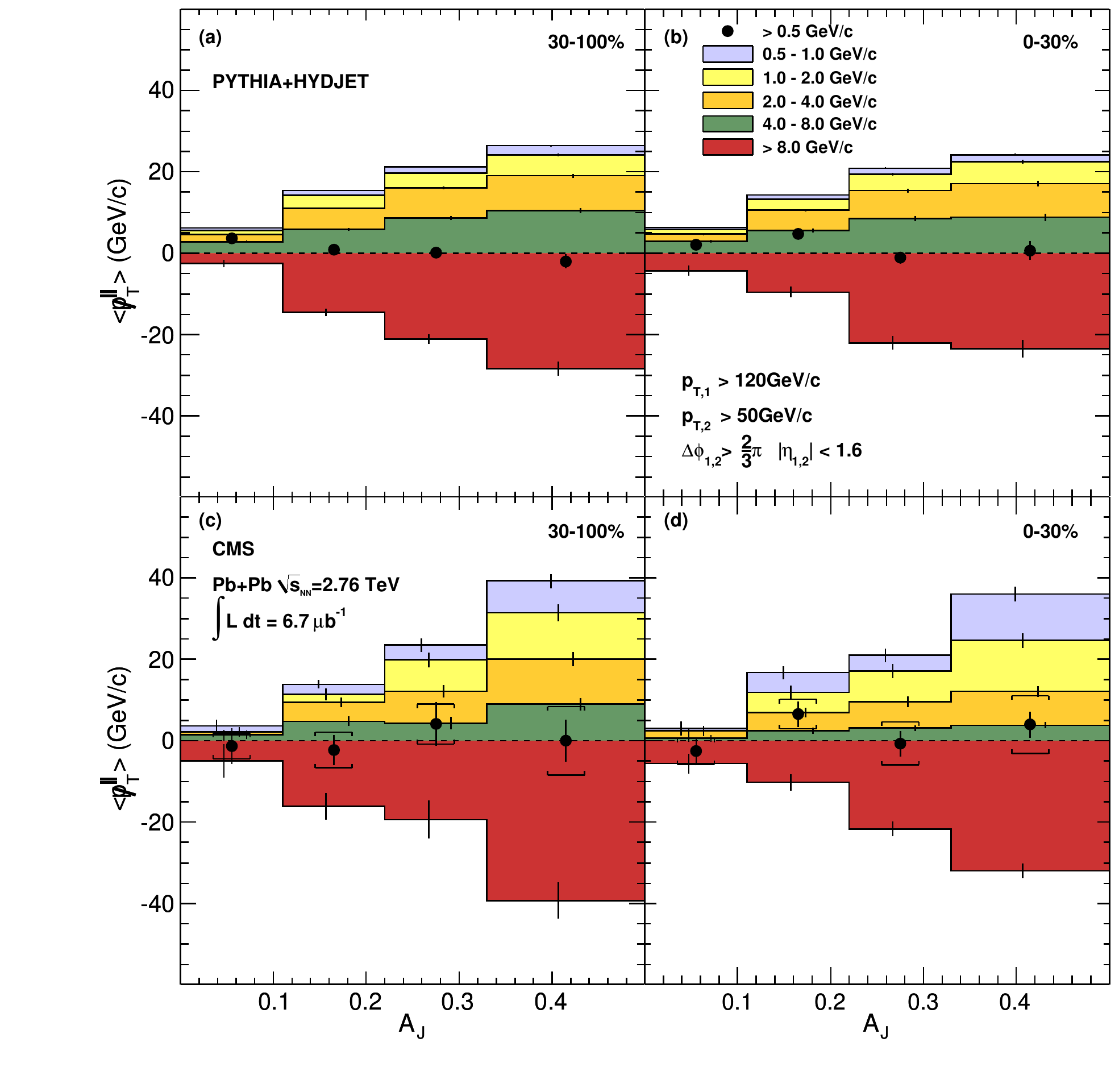}
    \caption{Average missing transverse momentum, 
$\langle \displaystyle{\not} p_{\mathrm{T}}^{\parallel} \rangle$, 
for tracks with $\pt > 0.5$~\GeVc, projected onto the leading jet axis (solid circles).
The $\langle \displaystyle{\not} p_{\mathrm{T}}^{\parallel} \rangle$ values are shown as a function of dijet asymmetry
$A_J$ for 30--100\% centrality (left) and 0--30\% centrality (right).
For the solid circles, vertical bars and brackets represent 
the statistical and systematic uncertainties, respectively.
Colored bands show the contribution to $\langle \displaystyle{\not} p_{\mathrm{T}}^{\parallel} \rangle$ for five
ranges of track \pt. The top and bottom rows show results for {\sc{pythia+hydjet}} and \PbPb\ data, respectively. 
For the individual $\pt$ ranges, the statistical uncertainties are shown as vertical bars.
}
    \label{fig:MissingpT}
  \end{center}
\end{figure}

The requirements of the background subtraction procedure limit the track-jet correlation study
to tracks with $p_{\mathrm{T}} > 1.0$~\GeVc\  and $\Delta R < 0.8$. Complementary information about the 
overall momentum balance in the dijet events can be obtained using the projection of missing 
\pt\ of reconstructed charged tracks onto the leading jet axis. For each event, this 
projection was calculated as 

\begin{equation}
\displaystyle{\not} p_{\mathrm{T}}^{\parallel} = 
\sum_{\rm i}{ -p_{\mathrm{T}}^{\rm i}\cos{(\phi_{\rm i}-\phi_{\rm Leading\ Jet})}},
\end{equation}
where the sum is over all tracks with $\pt > 0.5$~\GeVc\ and $|\eta| < 2.4$. The results were 
then averaged over events to obtain $\langle \displaystyle{\not} p_{\mathrm{T}}^{\parallel} \rangle$.
No background subtraction was applied, which allows this study to include the $|\eta_{jet}| < 0.8$ and 
$0.5 < p_{\mathrm{T}}^{\rm Track} < 1.0$~\GeVc\ regions not accessible for the study in Section~\ref{sec:track_jet_correlations}.
The leading and subleading jets were again required to have $|\eta| < 1.6$.

In Fig.~\ref{fig:MissingpT}, $\langle \displaystyle{\not} p_{\mathrm{T}}^{\parallel} \rangle$
is shown as a function of \AJ\ for two centrality bins, 30--100\% (left) and 0--30\% (right). 
Results for {\sc{pythia+hydjet}} are presented in the top row, while the bottom row shows the results
for \PbPb\ data.
Using tracks with $|\eta| < 2.4$ and $\pt > 0.5$~\GeVc, one sees
that indeed the momentum balance of the events, shown as solid circles, is recovered within uncertainties, 
for both centrality ranges and even for events with large observed dijet asymmetry, in both data and simulation. 
This shows that the dijet momentum imbalance is not related to undetected activity in the event due to instrumental 
(e.g. gaps or inefficiencies in the calorimeter) or physics (e.g. neutrino production) effects.

\begin{figure}[th!]
  \begin{center}
    \includegraphics[width=1.0\textwidth]{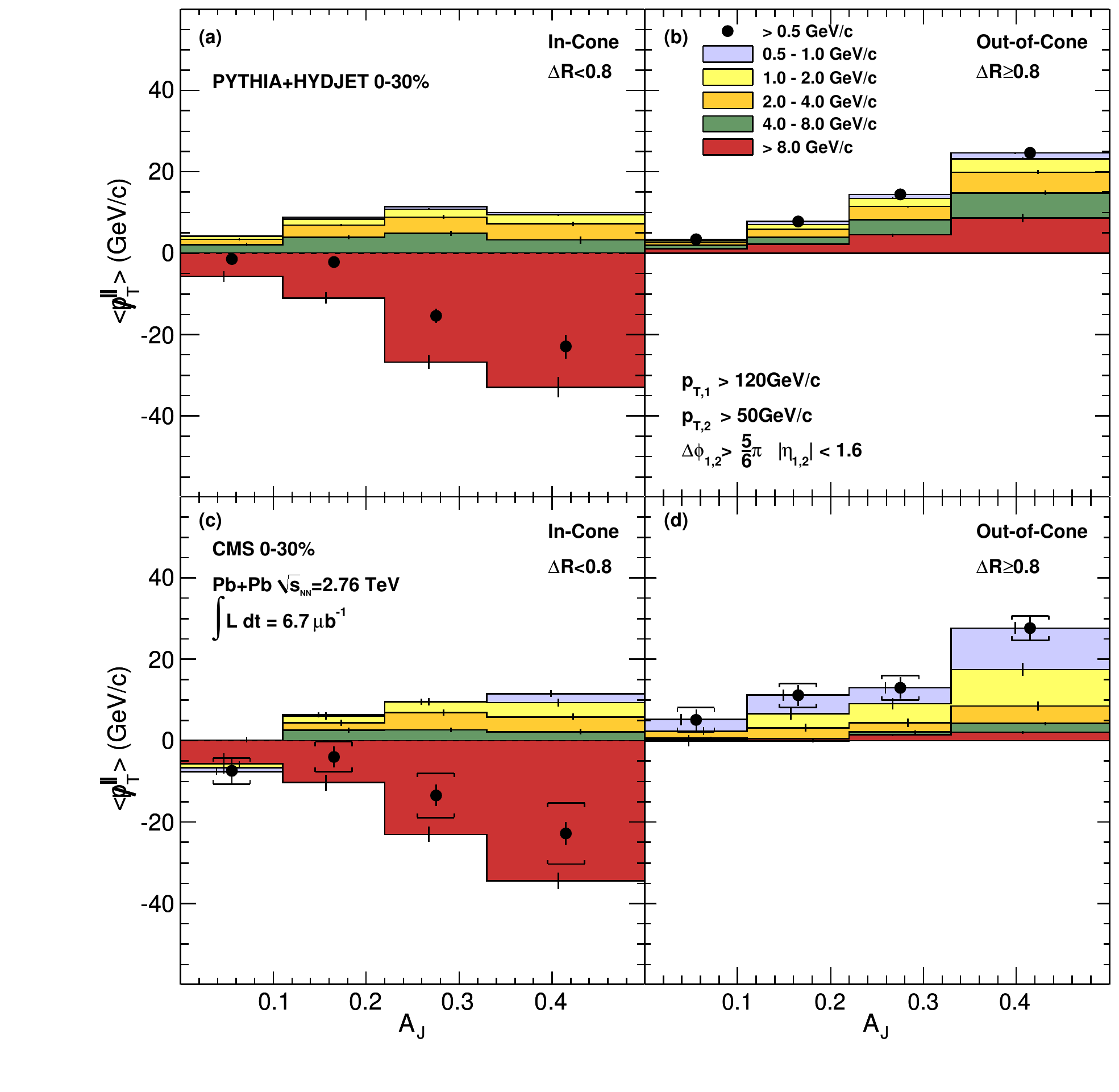}
    \caption{Average missing transverse momentum, 
$\langle \displaystyle{\not} p_{\mathrm{T}}^{\parallel} \rangle$, 
for tracks with $\pt > 0.5$\GeVc, projected onto the leading jet axis (solid circles).
The $\langle \displaystyle{\not} p_{\mathrm{T}}^{\parallel} \rangle$ values are shown as a function of dijet asymmetry
$A_J$ for 0--30\% centrality, inside ($\Delta R < 0.8$) one of the leading or subleading jet cones (left) and 
outside ($\Delta R > 0.8$) the leading and subleading jet cones (right).
For the solid circles, vertical bars and brackets represent 
the statistical and systematic uncertainties, respectively.
For the individual $\pt$ ranges, the statistical uncertainties are shown as vertical bars.  }
    \label{fig:MissingpTInConeOutCone}
  \end{center}
\end{figure}
The figure also shows the contributions to 
$\langle \displaystyle{\not} p_{\mathrm{T}}^{\parallel} \rangle$ for five transverse momentum ranges from 0.5--1~\GeVc\ to
$\pt > 8$~\GeVc. The vertical bars for each range denote statistical uncertainties. 
 For data and simulation, a large negative contribution to $\langle \displaystyle{\not} p_{\mathrm{T}}^{\parallel} \rangle$
(i.e., in the direction of the leading jet) by the $\pt > 8$~\GeVc\ range is balanced by the combined contributions from 
the  0.5--8~\GeVc\ regions.  Looking at the $\pt < 8$~\GeVc\ region in detail, important differences
between data and simulation emerge. For {\sc{pythia+hydjet}} both centrality ranges show a large 
balancing contribution from the intermediate \pt\ region of 4--8~\GeVc, while the contribution from
the two regions spanning 0.5--2~\GeVc\ is very small. In peripheral \PbPb\ data, the contribution of 
0.5--2~\GeVc\ tracks relative to that from 4--8~\GeVc tracks is somewhat enhanced compared to the simulation.
In central \PbPb\ events, the relative contribution of low and intermediate-\pt\  tracks
is actually the opposite of that seen in {\sc{pythia+hydjet}}. 
In data, the 4--8~\GeVc\ region makes almost no contribution to the overall
momentum balance, while a large fraction of the negative imbalance from high \pt\ is recovered in low-momentum tracks.

The dominant systematic uncertainty for the \pt\ balance measurement comes
from the \pt-dependent uncertainty in the track reconstruction efficiency
and fake rate described in Section~\ref{sec:track_jet_correlations}. 
A 20\% uncertainty was assigned to the final result, stemming from the residual difference
between the {\sc{pythia}} generator-level and the reconstructed {\sc{pythia+hydjet}} tracks
at high \pt. This is combined with an absolute 3~\GeVc\ uncertainty
that comes from the imperfect cancellation of the background tracks. The background effect
was cross-checked in data from a random cone study in 0--30\% central events
similar to the study described in Section~\ref{sec:track_jet_correlations}.
The overall systematic uncertainty is shown as brackets in Figs.~\ref{fig:MissingpT}
and \ref{fig:MissingpTInConeOutCone}.

Further insight into the radial dependence of the momentum balance can be gained by studying 
$\langle \displaystyle{\not} p_{\mathrm{T}}^{\parallel} \rangle$ separately for tracks inside cones of size $\Delta R = 0.8$
around the leading and subleading jet axes, and for tracks outside of these cones. The results of this study for central events
are shown in Fig.~\ref{fig:MissingpTInConeOutCone} for the in-cone balance and out-of-cone balance for MC and data.
As the underlying \PbPb\ event in both data and MC is not $\phi$-symmetric on an event-by-event basis, the back-to-back requirement
was tightened to $\dphi_{12} > 5 \pi/6$ for this study.

One observes that for both data and MC an in-cone imbalance of $\langle \displaystyle{\not} p_{\mathrm{T}}^{\parallel} \rangle \approx
-20$~\GeVc\ is found for the $\AJ > 0.33$ selection. In both cases this is balanced by a corresponding out-of-cone 
imbalance of  $\langle \displaystyle{\not} p_{\mathrm{T}}^{\parallel} \rangle \approx 20$~\GeVc. However, in the \PbPb\ data 
the out-of-cone contribution is carried almost entirely  by tracks with $0.5 < \pt < 4$~\GeVc\, whereas in MC more than 50\% of the balance 
is carried by tracks with $\pt > 4$~\GeVc, with a negligible contribution from $\pt < 1$~\GeVc.

The {\sc{pythia+hydjet}} results are indicative of semi-hard initial or final-state radiation as the underlying cause for large 
\AJ\ events in the MC study. This has been confirmed by further studies which showed that in {\sc{pythia}} the momentum balance 
in the transverse plane for events with large \AJ\ can be restored if a third jet with $\pt > 20$~\GeVc, which is present in more 
than 90\% of these events, is included. This is in contrast to the results for large-\AJ\ \PbPb\ data, which show that a
large part of the momentum balance is carried by soft particles ($\pt < 2$~\GeVc) and radiated
at large angles to the jet axes ($\Delta R > 0.8$).

\section{Summary}
\label{sec:summary}

The CMS detector has been used to study jet production in PbPb collisions at 
$\sqrt{s_{_{NN}}}=$\,2.76~TeV. Jets were reconstructed using
primarily the calorimeter information in a data sample corresponding to an integrated luminosity of
$L_{\rm int}=6.7 ~\mu$b$^{-1}$. Events having a leading jet with $\pt > 120$~\GeVc\ and 
$|\eta| < 2$ were selected. As a function of centrality, dijet events with a subleading jet 
of $\pt > 50$~\GeVc\ and $|\eta| < 2$ were found to have an
increasing momentum imbalance. Data were compared to {\sc{pythia}} dijet simulations for
\pp\ collisions at the same energy which were embedded into real heavy ion events.
The momentum imbalances observed in the data were significantly larger than those
predicted by the simulations.
While the relative imbalance between the 
leading and subleading jets increased with increasing collision centrality,
it was found to be largely independent of the leading jet \pt, up to the highest \pt region studied ($\approx$210~\GeVc). 

The angular distribution of jet fragmentation products has been explored
by associating charged tracks with the dijets observed in the calorimeters. 
The calorimeter-based momentum imbalance is reflected in the associated track 
distributions, which show a softening and widening of the subleading jet 
fragmentation pattern
for increasing dijet asymmetry, while the high-\pt components of the leading jet remain nearly
unchanged.

Studies of the missing transverse momentum projected on the jet axis have shown that the overall momentum 
balance can be recovered if tracks at low \pt\ are included. In the PbPb data, but not in the simulations, 
a large fraction of the balancing momentum is 
carried by tracks having  $\pt < 2$~\GeVc. 
Comparing the momentum balance inside and outside of cones of $\Delta R = 0.8$ 
around the leading and subleading jet axes demonstrates that a large contribution to 
the momentum balance in data arises from soft particles radiated at $\Delta R > 0.8$
to the jets, a feature which is also not reproduced in {\sc{pythia}} calculations.

In conclusion, a strong increase in the fraction of highly unbalanced jets has been
seen in central PbPb collisions compared with peripheral collisions and
model calculations, consistent with a high degree of jet quenching in the produced matter.  A large fraction of the momentum balance of these unbalanced jets is 
carried by low-\pt particles at large radial distance, in contrast
to {\sc{pythia}} simulations embedded into heavy ion events.
The results provide qualitative constraints on the nature of the jet modification 
in \PbPb\ collisions and quantitative input to models of the transport properties of the medium
created in these collisions.

\subsection*{Acknowledgments}
\hyphenation{Bundes-ministerium Forschungs-gemeinschaft Forschungs-zentren} We wish to congratulate our colleagues in the CERN accelerator departments for the excellent performance of the LHC machine. We thank the technical and administrative staff at CERN and other CMS institutes. This work was supported by the Austrian Federal Ministry of Science and Research; the Belgium Fonds de la Recherche Scientifique, and Fonds voor Wetenschappelijk Onderzoek; the Brazilian Funding Agencies (CNPq, CAPES, FAPERJ, and FAPESP); the Bulgarian Ministry of Education and Science; CERN; the Chinese Academy of Sciences, Ministry of Science and Technology, and National Natural Science Foundation of China; the Colombian Funding Agency (COLCIENCIAS); the Croatian Ministry of Science, Education and Sport; the Research Promotion Foundation, Cyprus; the Estonian Academy of Sciences and NICPB; the Academy of Finland, Finnish Ministry of Education, and Helsinki Institute of Physics; the Institut National de Physique Nucl\'eaire et de Physique des Particules~/~CNRS, and Commissariat \`a l'\'Energie Atomique et aux \'Energies Alternatives~/~CEA, France; the Bundesministerium f\"ur Bildung und Forschung, Deutsche Forschungsgemeinschaft, and Helmholtz-Gemeinschaft Deutscher Forschungszentren, Germany; the General Secretariat for Research and Technology, Greece; the National Scientific Research Foundation, and National Office for Research and Technology, Hungary; the Department of Atomic Energy, and Department of Science and Technology, India; the Institute for Studies in Theoretical Physics and Mathematics, Iran; the Science Foundation, Ireland; the Istituto Nazionale di Fisica Nucleare, Italy; the Korean Ministry of Education, Science and Technology and the World Class University program of NRF, Korea; the Lithuanian Academy of Sciences; the Mexican Funding Agencies (CINVESTAV, CONACYT, SEP, and UASLP-FAI); the Pakistan Atomic Energy Commission; the State Commission for Scientific Research, Poland; the Funda\c{c}\~ao para a Ci\^encia e a Tecnologia, Portugal; JINR (Armenia, Belarus, Georgia, Ukraine, Uzbekistan); the Ministry of Science and Technologies of the Russian Federation, and Russian Ministry of Atomic Energy; the Ministry of Science and Technological Development of Serbia; the Ministerio de Ciencia e Innovaci\'on, and Programa Consolider-Ingenio 2010, Spain; the Swiss Funding Agencies (ETH Board, ETH Zurich, PSI, SNF, UniZH, Canton Zurich, and SER); the National Science Council, Taipei; the Scientific and Technical Research Council of Turkey, and Turkish Atomic Energy Authority; the Science and Technology Facilities Council, UK; the US Department of Energy, and the US National Science Foundation.
Individuals have received support from the Marie-Curie programme and the European Research Council (European Union); the Leventis Foundation; the A. P. Sloan Foundation; the Alexander von Humboldt Foundation; the Associazione per lo Sviluppo Scientifico e Tecnologico del Piemonte (Italy); the Belgian Federal Science Policy Office; the Fonds pour la Formation \`a la Recherche dans l'Industrie et dans l'Agriculture (FRIA-Belgium); and the Agentschap voor Innovatie door Wetenschap en Technologie (IWT-Belgium).

\bibliography{auto_generated}   

\cleardoublepage\appendix\section{The CMS Collaboration \label{app:collab}}\begin{sloppypar}\hyphenpenalty=5000\widowpenalty=500\clubpenalty=5000\textbf{Yerevan Physics Institute,  Yerevan,  Armenia}\\*[0pt]
S.~Chatrchyan, V.~Khachatryan, A.M.~Sirunyan, A.~Tumasyan
\vskip\cmsinstskip
\textbf{Institut f\"{u}r Hochenergiephysik der OeAW,  Wien,  Austria}\\*[0pt]
W.~Adam, T.~Bergauer, M.~Dragicevic, J.~Er\"{o}, C.~Fabjan, M.~Friedl, R.~Fr\"{u}hwirth, V.M.~Ghete, J.~Hammer\cmsAuthorMark{1}, S.~H\"{a}nsel, C.~Hartl, M.~Hoch, N.~H\"{o}rmann, J.~Hrubec, M.~Jeitler, G.~Kasieczka, W.~Kiesenhofer, M.~Krammer, D.~Liko, I.~Mikulec, M.~Pernicka, H.~Rohringer, R.~Sch\"{o}fbeck, J.~Strauss, F.~Teischinger, P.~Wagner, W.~Waltenberger, G.~Walzel, E.~Widl, C.-E.~Wulz
% ...
\vskip\cmsinstskip
% ...
1:~~Also at CERN, European Organization for Nuclear Research, Geneva, Switzerland\\
% ...

\end{sloppypar}
\end{document}